\documentclass[12pt]{article}
\usepackage[T1]{fontenc} 
\fontfamily{bch}\selectfont
\usepackage[skip=8pt plus1pt, indent=0pt]{parskip}
\usepackage{titlesec}
\usepackage{fancyhdr}
\usepackage{setspace}
\usepackage{lmodern}
\usepackage[hypcap=false]{caption}
\usepackage{subcaption}
\usepackage[none]{hyphenat}

\newlength\titleindent
\newlength\sectionskip
\titleformat{\section}  
  {\vspace{\sectionskip}\fontsize{24}{24}\sffamily\raggedright} 
  {\llap{\parbox[b]{\titleindent}{\thesection\hfill}}} 
  {0em} 
  {} 
  [] 
\titlespacing*{\section}{0pt}{0pt}{24pt plus 1pt}

\titleformat{\subsection}  
  {\fontsize{15}{15}\sffamily\bfseries\raggedright} 
  {\thesubsection} 
  {0.6em} 
  {} 
  [] 
\titlespacing*{\subsection}{0pt}{18pt plus 3pt minus 1pt}{10pt plus 1pt}

\titleformat{\subsubsection}  
  {\fontsize{13}{13}\sffamily\bfseries\raggedright} 
  {\thesubsubsection} 
  {0.6em} 
  {} 
  [] 
\titlespacing*{\subsection}{0pt}{24pt plus 3pt minus 1pt}{12pt plus 1pt}

\usepackage[headheight=15pt]{geometry}  
\usepackage{mathrsfs}                   
\usepackage{graphicx}                   
\graphicspath{ {figures/} }             
\usepackage{booktabs}                   
\usepackage[table,xcdraw]{xcolor}       
\usepackage{multirow}                   
\usepackage[normalem]{ulem}             
\useunder{\uline}{\ul}{}                
\usepackage{appendix}                   
\usepackage{xspace}
\usepackage{amsthm}                     
\usepackage{hyperref}                   
\usepackage{natbib}                     
\usepackage{lscape}                     
\usepackage{mathtools}                  
\usepackage{algpseudocode}              
\usepackage[Algorithmus]{algorithm}     
\usepackage{tabularx}                   

\hypersetup{
    pdftitle={Model-based traffic state estimation using camera-equipped probe vehicles},
    pdfauthor={Tanay Rastogi},
    colorlinks=true,
    linkcolor=black,
    citecolor=black,
    filecolor=black,
    urlcolor=black
}

\begin{document}

\title{Model-based traffic state estimation using camera-equipped probe vehicles}
\author{Tanay Rastogi, Michele D. Simoni and Anders Karlström}
\date{}
\maketitle
\pagenumbering{arabic}

\begin{abstract}
\noindent This study addresses the challenge of estimating traffic states for road links. We propose an innovative approach that leverages partial trajectory data captured by camera-equipped probe vehicles traveling in the opposite lane. The methodology combines state-of-the-art computer vision algorithms for extracting vehicle trajectories from street-view video sequences with a novel estimation technique based on the Cell Transmission Model (CTM) and Genetic Algorithms (GA). Our approach first calibrates Fundamental Diagram (FD) parameters using observed cell densities, then estimates boundary conditions for all space-time diagrams. We validate the method using simulated traffic data from three different types of links and parameter settings. Results show that the proposed methodology can estimate traffic densities in unobserved regions, even with limited data availability. This research contributes to the field by introducing a cost-effective, high-resolution traffic data collection method and a robust estimation technique for comprehensive traffic state information. While the study shows promising results, it also identifies areas for improvement, including refining models, optimizing processes, and testing with real-world data to enhance accuracy and scalability.
\end{abstract}

\textbf{Keywords}: traffic state estimation,  genetic algorithm, cell transmission model

\newpage
\section{Introduction}
\label{Introduction}
The term "traffic state" typically refers to set of variables that quantitively describe the condition of traffic, including density, speed and flow. To effectively implement intelligent traffic management strategies and operations, it is crucial to accurately measure traffic conditions on the road network. These approaches are based on real-time data collection from large-scale road networks. The conventional method to collect traffic data is using stationary sensors, such as loop detectors, surveillance cameras, and radar-based sensors. These sensors provide traffic state measurement for a specific fixed location with high temporal resolution but limited spatial coverage. Many studies have used the data from these sensors to estimate travel time (\cite{Robinson2005, Li2018}), speed (\cite{Coifman2009}), traffic densities (\cite{Panda2019, Timotheou2015}), traffic jam (\cite{Tyagi2012}, and other traffic characteristics.

\begin{figure}[htbp]
\centering
{\resizebox*{10cm}{!}{\includegraphics{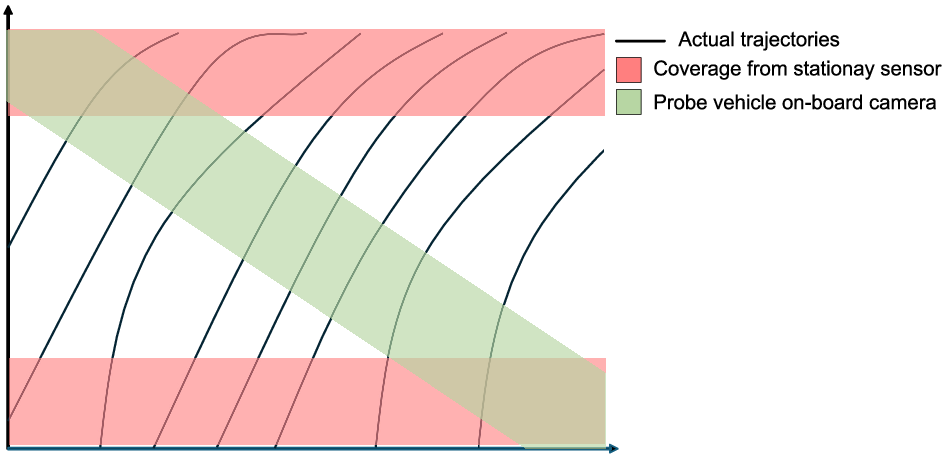}}}
\caption{Illustration of the space-time diagram highlighting sections measured by stationary, represented in RED and onboard camera sensors, represented in GREEN.}
\label{fig:coverage}
\end{figure}

Recently, lightweight portable sensors such as smartphones, dashboard cameras (dash-cams), and embedded devices have become popular and can be mounted on vehicles to capture traffic incidents. The progress of in-vehicle camera technology has opened up significant possibilities for traffic data collection using street-view video sequences, both at macroscopic and microscopic levels. By employing state-of-the-art computer vision algorithms and GPS measurements, these video sequences allow for the extraction of vehicle trajectories, which can be used to generate space-time diagrams at the link level. Compared to stationary sensors, on-board vehicle cameras provide higher spatial resolution for traffic state information along the entire link. (\cite{Cao2011}). However, neither stationary sensors nor on-board vehicle cameras can continuously collect data over an entire link for extended periods. As illustrated in Figure \ref{fig:coverage}, there remains a considerable portion of the space-time diagram that is unobserved by both sensor types. To address this limitation, traffic states in unobserved regions must be inferred using partially observed data through Traffic State Estimation (TSE) methods. TSE involves inferring the flow, density, and speed of road segments based on partial observations and prior knowledge of traffic patterns.

In this context, this study introduces an innovative approach to estimate traffic states within unobserved regions of a space-time diagram for a specific road link, leveraging high-resolution data collected from on-board vehicle cameras. We propose a novel method that employs the Cell Transmission Model (CTM) to simulate traffic dynamics and a Genetic Algorithm (GA) to calibrate the Fundamental Diagram (FD) parameters and boundary conditions necessary for estimating traffic densities. A key aspect of this methodology is that it operates offline, processing the collected data only after a sufficient number of probe vehicle runs have been completed for the target link.

The main contributions of this study are summarized below.
\begin{itemize}
    \item Demonstration of data collection using onboard cameras for high-resolution traffic data collection along entire road links.
    \item Introduction of a novel approach for traffic state estimation using vehicle trajectories obtained from cameras mounted on moving vehicles in the opposite lane.
    \item Proposal of an innovative estimation method that uses the CTM in conjunction with a GA to calibrate FD parameters and boundary conditions for traffic state estimation.
    \item Experiments from different types of link and parameter setting using the simulated traffic data to demonstrate the proposed method's effectiveness at estimation.
\end{itemize}

The paper is structured as follows: Section \ref{sec:references} begins with a review of the existing literature on the topic, identifying the research gaps that motivate this study. In Section \ref{sec:kitti}, we detail the method for collecting vehicle trajectories using camera-equipped probe vehicles travelling in the opposite lane to observe forward-moving traffic. Following this, Section \ref{sec:stateEstimation} introduces the proposed methodology for estimating traffic states based on the collected on-board camera data. Due to the lack of real-world data, we validate the proposed method using simulated data from a traffic simulation of a city, designed to represent actual traffic conditions. The experimental setup, results, and discussion are provided in Section \ref{sec:experiments}. Finally, Section \ref{sec:conclusion} concludes the paper by summarizing the analysis and findings and highlighting the potential of this innovative approach for traffic state estimation using on-board camera data.

\section{Related Works}\label{sec:references}
Space-time trajectory diagrams are essential in traffic engineering and transportation planning because they visually represent vehicle movements over time and space, offering crucial insights into traffic flow dynamics. Traditionally, data for these diagrams has been collected using Floating Car Data (FCD), which involves mobile sensors installed in vehicles during their journeys. Numerous studies have utilized space-time diagrams derived from FCD to estimate traffic states. For instance, \cite{Herrera2010} integrated GPS data into traffic flow models for TSE. \cite{Nantes2016} developed a methodology for traffic prediction by combining data from loop detectors with partial observations from Bluetooth and GPS devices. \cite{Chaturvedi2021} introduced a method for estimating traffic conditions using DSRC-based mobile sensors. \cite{Han2013} proposed a methodology to extract space-time traffic patterns across large-scale transportation networks. While several studies have employed cameras for TSE, most have focused on data collected from fixed cameras, such as high-mounted surveillance or highway cameras. For example, \cite{Pletzer2012, Li2013, Ua-Areemitr2019} utilized traffic counts from static cameras to estimate traffic states and assess service levels for specific road sections visible to these cameras.

There is a noticeable gap in research regarding the use of on-board vehicle cameras as a primary data source for traffic flow data collection. Recently, \cite{Seo2015PV} created a way to estimate traffic flow variables using probe vehicles equipped with equipment for measuring spacing. This method does not require any external assumptions, such as a fundamental diagram. Research focusing on collecting traffic data from the opposite lane using on-board cameras is even scarcer. Notable examples include \cite{Guerrieri2021}, who proposed an automatic traffic data acquisition method using deep learning to detect vehicles in the opposite lane and estimate macroscopic traffic variables such as flow rate, space mean speed, and vehicle density. Similarly, \cite{Kumar2021} introduced a novel algorithm to estimate citywide cross-sectional traffic flow using data collected in the opposite lane from moving camera videos by utilizing deep learning-based techniques for vehicle detection, tracking, and photogrammetry. 

While these studies serve as valuable references, our research uniquely focuses on partial trajectories of vehicles in the opposite lane, collected by an on-board camera-equipped backward probe. Unlike previous studies that consider traffic on the same link and use headway spacing as input, we focus on collecting vehicle trajectories in the opposite lane.

Current research on TSE categorizes the process into three fundamental components: input data, dynamic traffic flow models, and estimation methods. The input data in TSE refers to partial observations obtained from the road network, which are used to estimate unobserved traffic states. The traffic flow model provides a structured framework for capturing the dynamics of traffic flow, while estimation methods encompass various techniques used to infer traffic conditions based on limited observations and the pre-existing knowledge of the traffic model \cite{Seo2017}. 

In our study, the space-time diagrams generated from vehicles detected by probe cameras in the opposite lane constitute the input data for TSE. These extracted trajectories provide partial information for the road link and form the basis for estimating unobserved regions of the diagram. One of the most widely used traffic flow models is the Cell Transmission Model (CTM), first introduced by \cite{Daganzo1994}. Later \cite{Munoz2004} introduced a modified version of CTM that uses cell densities as state variables instead of cell occupancy and accommodates non-uniform cell lengths. This modified CTM accurately reproduces observed freeway traffic behaviour, including bottleneck formation and the spatial and temporal extent of congestion. For example, CTM has been used to model hard shoulder running operations with queue warning services during one-time traffic accidents \cite{Li2017}, and it has also been used with deep reinforcement learning to improve ramp metering performance on freeways when there are uncertain bottleneck conditions \cite{Zheng2024}. Numerous studies have used variations of CTM to estimate traffic states based on trajectory data. For example, \cite{Seo2015TSE} utilized ensemble learning and a traffic flow model to estimate traffic states based on data collected by advanced probe vehicles equipped with spacing measurement tools, mitigating the negative effects of high fluctuations in microscopic vehicular traffic. \cite{Takenouchi2019} suggested a variational theory-based way to estimate traffic conditions using measurements from a vehicle going the opposite way (the backward probe vehicle) along with data from a regular probe vehicle. They also looked at how sensitive estimates were to changes in the input data and measurements. \cite{Kuwahara2021} extended this research by estimating cell density time-dependently using a State Space Model with CTM to estimate traffic states under incident conditions on an expressway section, using measurements from a combination of backward-moving and forward-moving probe vehicles, along with traffic detectors. In addition, \cite{Fulari2016} described a dynamical systems approach that uses Kalman filtering and CTM to estimate traffic speed and density in real time by using data fusion where location-based and spatial traffic variables were used from automated sensors. The error statistics of the automated sensor data were explicitly included to improve the accuracy of the estimates.

Previous studies typically assume that vehicle trajectories used as input data are complete and fully observed by the backward probe vehicle. However, this assumption often does not hold true in real-world scenarios. In light of this, we propose a novel estimation method that accounts for incomplete observations when estimating traffic states.

\section{Vehicle Trajectories Collection Method} \label{sec:kitti}
\begin{figure}[htbp]
\centering
{\resizebox*{11cm}{!}{\includegraphics{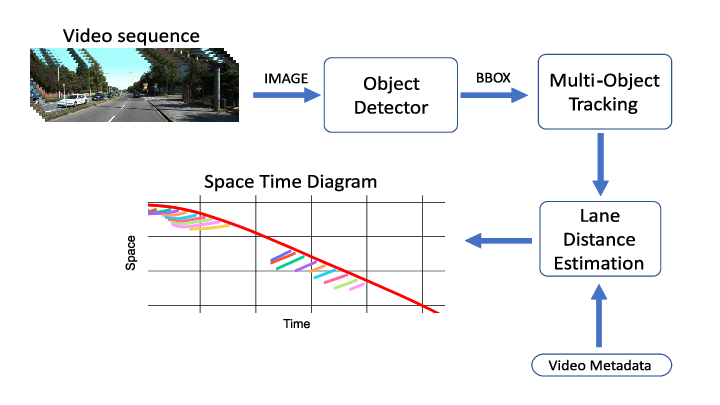}}}
\caption{The flowchart illustrates the process of extracting the space-time diagram from street-view video using the proposed methodology, as proposed by  (\cite{Tanay2023}).}
\label{fig:pipeline}
\end{figure}

In this study, we infer vehicle trajectories from street-view video sequences captured by camera-equipped probe vehicles on the opposite lane to observe forward-moving traffic. We employed state-of-the-art computer vision algorithms based on deep neural networks (DNN), photogrammetry, and geodesy to infer these trajectories. The inferred trajectories are used to construct time-space diagrams through three key steps: multi-object detection, multi-object tracking, and lane distance estimation, as illustrated in Figure \ref{fig:pipeline}. For detecting and identifying vehicles in each frame of the video sequence, we used the YOLOv5 multi-object detector. Subsequently, each detected vehicle is assigned a unique ID and tracked across consecutive frames using a multi-object tracker known as StrongSORT. Once vehicles have been detected and labelled, the distance of each detected vehicle on a road network is calculated using time-stamped GPS information, photogrammetry, and geodesy. Finally, utilizing the distance and timestamp information, the time-space diagram is generated, with the spatial dimensions corresponding to the link length and the time dimension representing the travel time of the camera-mounted vehicle. The details on training, implementation, and validation of each step of the proposed methodology are presented in the article by \cite{Tanay2023}.

\begin{landscape}
    \begin{figure}
    \centering
    \begin{subfigure}[b]{0.65\textwidth}
       \includegraphics[width=1\linewidth]{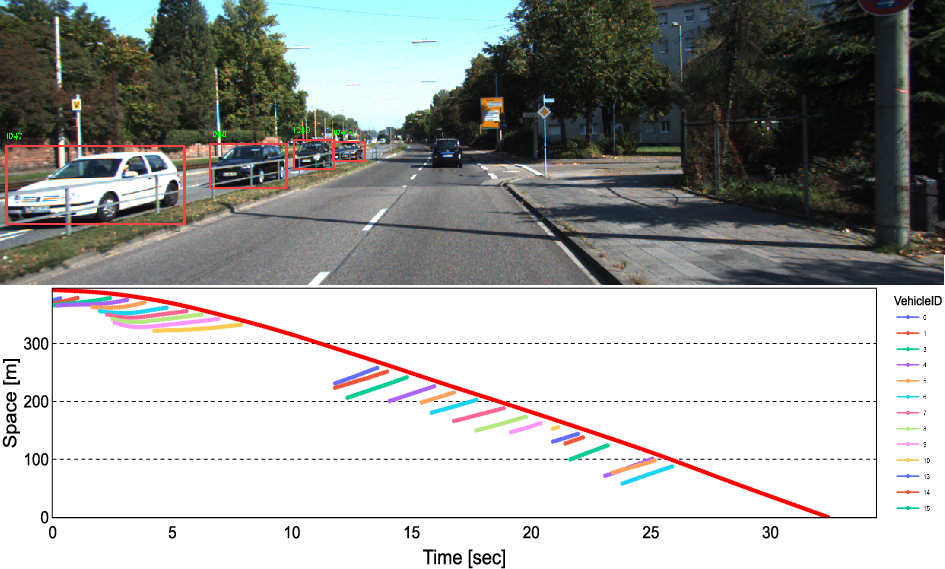}
       \caption{video-0004}
    \end{subfigure}
    
    \begin{subfigure}[b]{0.65\textwidth}
       \includegraphics[width=1\linewidth]{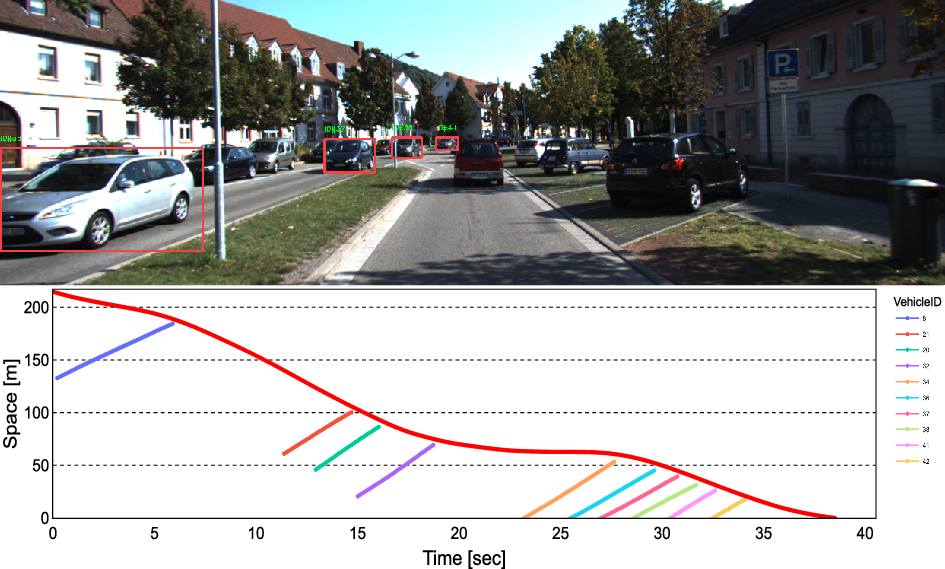}
       \caption{video-0011}
    \end{subfigure}
    
    \caption{Examples of videos from the KITTI dataset to illustrate the space-time diagram, extracted using the proposed methodology.}
    \label{fig:video}
    \end{figure}
\end{landscape}

To demonstrate the above-described data collection methodology, we utilized video sequences from an open-source computer vision dataset called KITTI (\cite{Geiger2012}). This dataset provides annotated street-view videos captured by a probe vehicle driving through urban environments, synchronized with GPS data. Figure \ref{fig:video} presents examples of two videos from the KITTI dataset, where the space-time diagram was extracted using the previously described methodology. In both videos, the RED coloured trajectory represents the probe vehicle, and the other trajectories represent vehicles in the opposite lane extracted using the proposed methodology.

The length of the extracted trajectory primarily depends on two factors: the detection capability of the object detector and the visibility of the detected vehicle on the camera. The YOLOv5 detector is a state-of-the-art object detector and performs well on the KITTI video sequences, as demonstrated in \cite{Tanay2023}. However, vehicles that are far from the camera appear too small for the object detector to accurately detect. Therefore, there is a limit to the maximum distance at which we can extract vehicle trajectories in the opposite lane. Additionally, trajectory information is lost once the detected vehicle in the opposite lane passes the probe vehicle. As a result, the space-time diagram contains information for only a small section, leaving the remaining portion unobserved. For the KITTI videos, we use the annotation data in the dataset to calculate the farthest distance at which objects are detected, and we set that as the limit for maximum camera detection distance. This is termed the camera's field of view ($cfv$) and is measured at \textbf{146 meters} for the KITTI dataset.

Figure \ref{fig:aggregate} illustrates camera detection distance as a RED-shaded region for the \textit{video-011}. Shaded portions mark the visible area within the camera, while the unshaded areas represent the unobserved regions in the space-time diagram.

\newpage

\subsection{Discretization}\label{sec:aggregation}
To calculate traffic states from the extracted trajectories, we discretized the space-time diagram into multiple smaller regions. Traffic states are estimated for each of these discretized space-time regions. The space-time diagram generated from video sequences has dimensions $L$ representing the length of the link and $T$ representing the total time taken by the probe vehicle to traverse that link. We divide the diagram into smaller cells with pre-determined spatial dimension $\Delta x$ and temporal dimension $\Delta t$.

Similar to the method used by \cite{Seo2015PV} and \cite{Chaturvedi2021}, we use traffic flow characteristics as defined by \cite{Edie1963}. According to Edie, the density $k_i(\mathbf{A})$ of cell $i$ in a space-time region $\mathbf{A_i}$ is given by:
\begin{align}
\label{eq:density}
k_i(\mathbf{A_i}) = \frac{\sum_{n\in N(\mathbf{A_i})} t (\mathbf{A_i})}{\left | \mathbf{A_i} \right |}
\end{align}

where $N(\mathbf{A_i})$ represents the number of vehicle trajectories and $t (\mathbf{A_i})$ is the time spent by all vehicles in the cell $i$. We calculate the $\mathbf{A_i}$ as the proportion of area of the cell that is visible in the camera. The value of $\mathbf{A}$ is given as, 
\begin{align}
\label{eq:area}
\left | \mathbf{A_i} \right | = \delta_i *(\Delta x * \Delta t)
\end{align}

where $\delta_i$ represents the portion of the cell $i$ that is in the camera's visible range. In this study, to ensure legible values during discretization, we only consider cells that have more than $50\%$ of their area visible on camera.

\begin{figure}[htbp]
\centering
{\resizebox*{15cm}{!}{\includegraphics{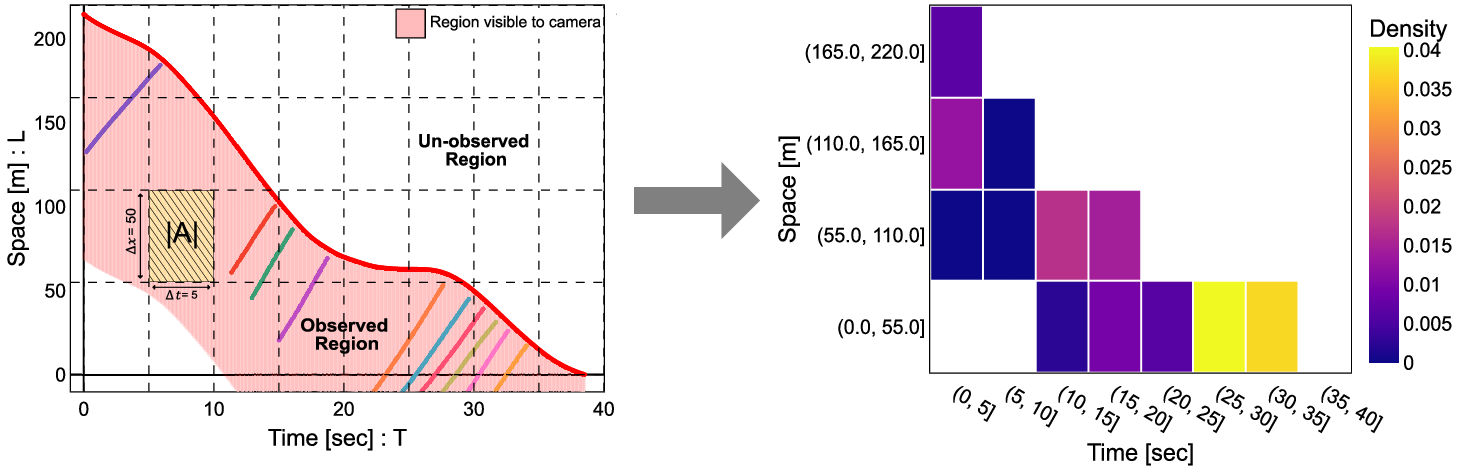}}}
\caption{Illustration to show the discretization of the trajectories in space-time diagram in KITTI-video0011 that are within the camera's visibility area. The portion that is within the RED-shaded region in the space-time diagram is discretized.} \label{fig:aggregate}
\end{figure}

To illustrate the discretization process, we use video-0011 as an example, which has a link length ($L$) of 200 meters, a total time ($T$) of 38 seconds, and cell dimensions of $\Delta x$ = 50 meters and $\Delta t$ = 5 seconds. Figure \ref{fig:aggregate} illustrates the aggregation of the space-time trajectory diagram into a density matrix. The cells in this matrix contain density information indicating the observed traffic states from the visible vehicle trajectories. As previously stated, we only consider cells with more than 50\% of their area visible to the camera. Consequently, the cell at the time interval $(5,10]$ seconds and space interval $(0,50]$ meters is excluded and marked as an unobserved region.

\section{Traffic State Estimation} \label{sec:stateEstimation}
To estimate the density of unobserved regions in the space-time diagram, we employ the CTM method, which requires a calibrated Fundamental Diagram (FD) and boundary conditions of the link. Our approach involves a two-step process using Genetic Algorithms (GA):
\begin{enumerate}
    \item We first calibrate the FD parameters using a GA, taking into account the cell densities in the observed cells.
    \item Once optimal FD values are determined, we apply another variation of GA-CTM to estimate the boundary conditions for all space-time diagrams.
\end{enumerate}
The final output of our proposed method is a complete space-time diagram where the unobserved cells are estimated using the optimal FD values and boundary conditions.

In our methodology, we utilize the CTM proposed by \cite{Munoz2004}. We acknowledge that this version of CTM assumes a steady-state speed-density relationship, which does not account for fluctuations around the equilibrium fundamental flow-density diagram. While several improvements have been proposed over the years, such as the Switching Mode Model (\cite{Munoz2006}), the Asymmetric CTM (\cite{Gomes2006}), the Enhanced Lagged CTM (\cite{Szeto2008}), and the Stochastic CTM (\cite{Sumalee2011}), we opt for the CTM proposed by Mu{\~{n}}oz, due to its computational efficiency and ability to capture important traffic phenomena while remaining analytically tractable. This choice allows us to balance accuracy with computational feasibility in our estimation process.

The specific implementation of CTM and GA used in our methodology is described in Sections \ref{sec:CTM} - \ref{sec:GA}, followed by detailed explanations of each step in the proposed estimation approach in Sections \ref{sec:FD_estimation}- \ref{sec:Density_Estimation}.

\begin{figure}[htbp]
\centering
{\resizebox*{15cm}{!}{\includegraphics{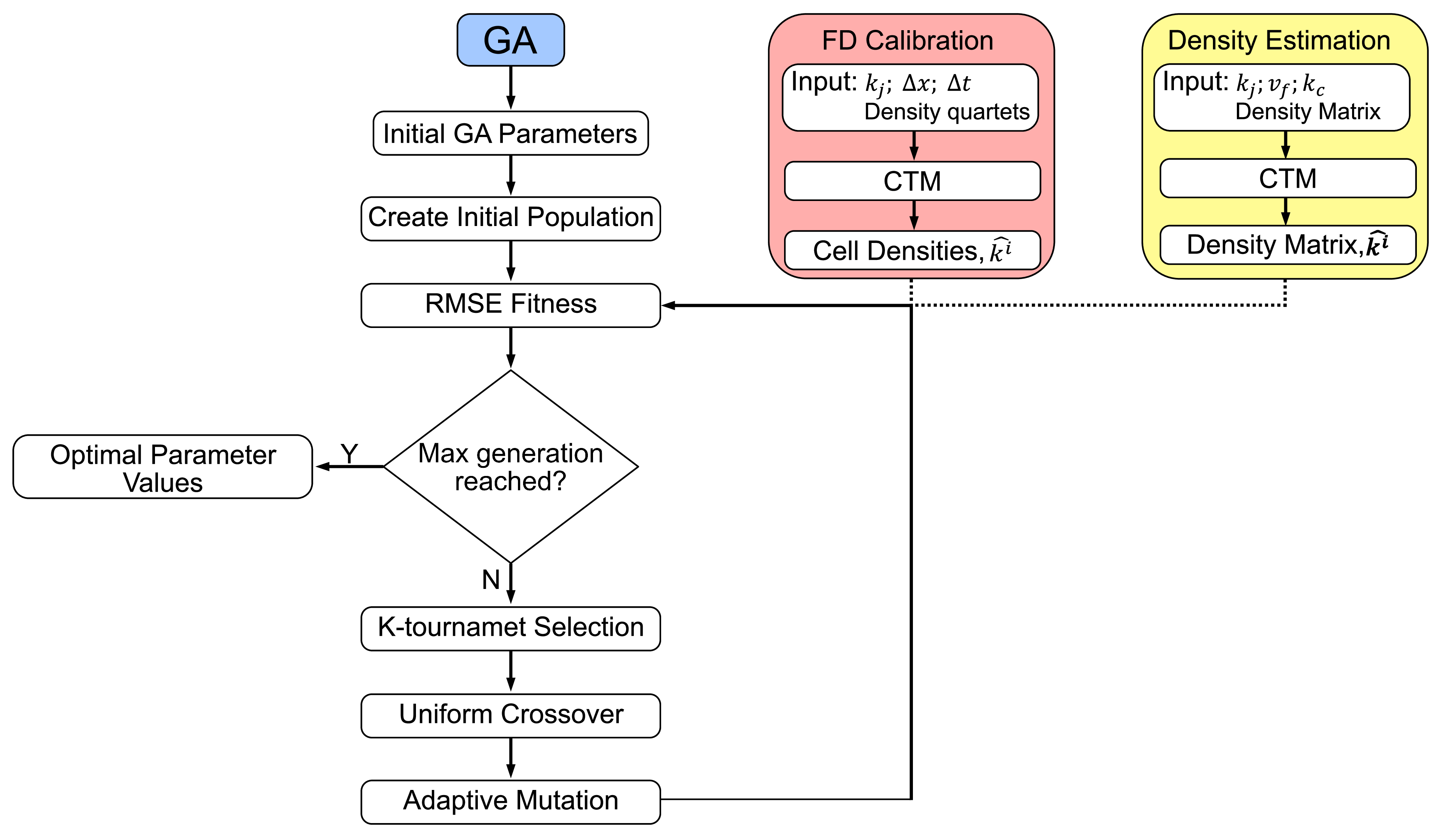}}}
\caption{The proposed GA method's flow chart illustrates the FD calibration and density estimation using CTM for fitness evaluation.} \label{fig:GA_flow}
\end{figure}

\subsection{Cell Transmission Model (CTM)}\label{sec:CTM}
The proposed methodology is based on CTM which describes the evolution of traffic density over time and space. According to CTM, the density of a cell $k$ at position $x$ and time $t + \Delta t$ depends on three factors: density of the cell at the previous time step $k(x,t)$; vehicle flow into the cell $Q(k(x,t),t)$ and vehicle flow out of the cell $Q(k(x+\Delta x, t), t)$. This relationship can be expressed mathematically as:
\begin{align}
\label{eq:CTM_single}
k(x, t+\Delta t)=k(x,t)+ \frac{\Delta t}{\Delta x}\left ( Q(k(x,t),t)-Q(k(x+\Delta x, t), t) \right )
\end{align} 

The flow in CTM $Q(k(x,t),t)$ is determined by the link's chosen Fundamental Diagram (FD). Assuming a triangular flow-density relationship, the flow is calculated as:
\begin{subequations}
\label{eq:tri_fd}
\begin{equation}
    Q(k(x,t),t) = \min\left \{ k(x-\Delta x,t) \cdot v_f \;; w_c \cdot (k_j - k(x,t))  \right \} \label{eq:trig_flow}
\end{equation}
\begin{equation}
    w_c = \frac{v_f \cdot k_c}{k_j-k_c} \label{eq:backwardSpeed}
\end{equation}
\end{subequations}
where $v_f$ is the free flow speed, $k_c$ is the optimal density, $k_j$ is the jam density and $w_{c}$ is the backward wave speed. The value of jam density $k_j$ can be calculated as the inverse of the minimum distance headway, $d$ and number of lanes, $NL$ for all vehicles on the link. The value is given as: 
\begin{align}
\label{eq:jam_density}
k_j = \frac{NL*1000}{d}
\end{align}

\subsection{Genetic Algorithm (GA)}\label{sec:GA}
The GA, created by \cite{JohnHolland1992}, is a computational optimization technique based on natural selection and evolution, initializing a population and iteratively evolving it through fitness calculation, selection, crossover, and mutation until the best solution is found or the maximum number of generations is reached. In our proposed methodology, we have customized the original GA to determine the optimal parameters for FD and boundary conditions. The flow chart illustrating the utilization of the GA in our methodology is presented in Figure \ref{fig:GA_flow}.

\newpage
The GA algorithm begins with randomly initialized candidate solutions for FD parameters or boundary conditions. These candidates represent potential values for either the FD parameters or boundary conditions, serving as potential solutions to the optimization problem. Each candidate's fitness is evaluated using a fitness function, which notably incorporates the CTM. The specific fitness calculations for FD parameter calibration and density estimation are explained in Section \ref{sec:FD_estimation} and Section \ref{sec:Density_Estimation}, respectively. The next step involves selecting the fittest individual using the K-way tournament selection procedure. These selected candidate solutions are then combined using a uniform crossover algorithm to create a new population for the next generation. Finally, an adaptive mutation method is applied, with mutation rates varying based on the fitness values of the candidates. This process repeats for subsequent generations.

\subsection{FD Calibration}\label{sec:FD_estimation}
To estimate the optimal parameter values for the Triangular FD, we utilized the cell densities from all discretized space-time diagrams. According to the CTM formulation in Eq.\ref{eq:CTM_single} and Eq.\ref{eq:tri_fd}, the cell density $k(x, t+\Delta t)$ depends on the values of $k(x,t)$, $k(x-\Delta x,t)$, $k(x+\Delta x,t)$, and the FD parameters $v_f$, $k_c$, and $k_j$. Given the other values in the CTM formulation, we can determine the values of $v_f$ and $k_c$ that best satisfy the relationship.

\begin{figure}[htbp]
\centering
{\resizebox*{10cm}{!}{\includegraphics{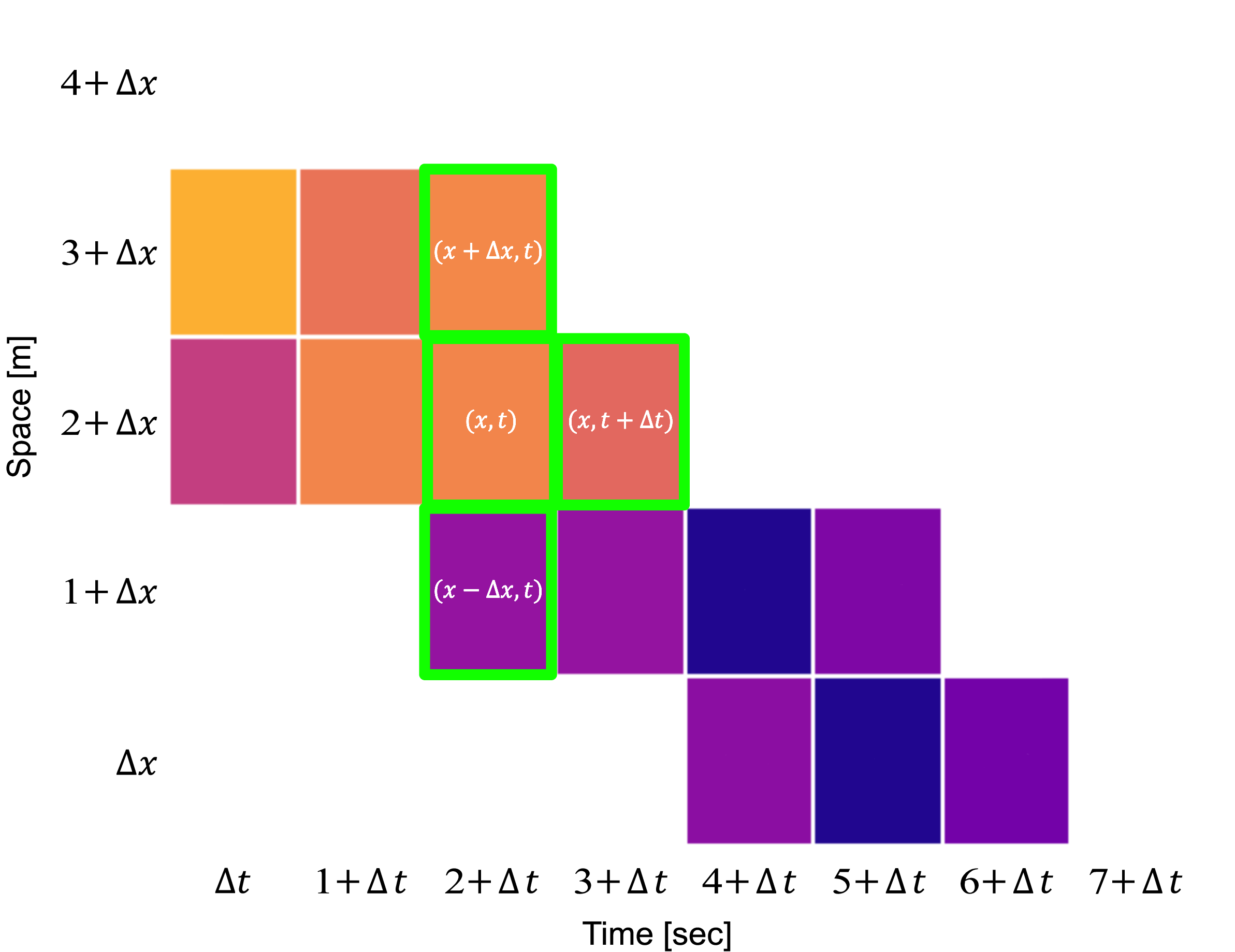}}}
\caption{Example of a space-time matrix where cells with a "green" boundary represent a \textbf{quartets} that is used to calibrate the triangular FD.}\label{fig:densityTuple}
\end{figure}

To achieve this, we extract the cell densities in the following exact formation, called \textbf{quartets}: $k(x,t)$, $k(x-\Delta x,t)$, $k(x+\Delta x,t)$, and $k(x, t+\Delta t)$ from all space-time density matrices. Figure \ref{fig:densityTuple} shows an example of a space-time density matrix where the cells marked with a green boundary are the cell densities used for the FD estimation.

We utilize the GA method, as described in Section \ref{sec:GA}, to determine the optimal values of $v_f$ and $k_c$ that satisfy Eq.\ref{eq:CTM_single} for all quartets of cell densities. To initialize the GA, the population is generated by pairing sampled values of $v_f$ and $k_c$ from a uniform distribution, considering the following constraints:

\begin{itemize}
    \item The value of $v_f$ is limited by the Courant-Friedrichs-Lewy (CFL) condition, which states: 
    \begin{equation}
        \label{eq:cfl}
        v_f \leq \frac{\Delta x}{\Delta t}
    \end{equation}
    \item The value of $k_c$ is constrained to always be less than half the jam density $k_j$. This constraint arises from the observation that congestion propagates through the mainstream at a slower speed than the maximum traffic speed. Therefore, the value of $k_c$ is given as: 
    \begin{equation}
        \label{eq:optimDensityConstrain}
        w_c < v_f \Rightarrow \frac{k_c}{k_j-k_c}< 1\Rightarrow k_c< \frac{k_j}{2}
    \end{equation}
\end{itemize}

We utilized Root Mean Square Error (RMSE) for the FD calibration step to determine the fitness of each possible solution in the GA process. The fitness of the $i^{th}$ pair of $v_f$ and $k_c$ in GA is calculated as the RMSE value between the actual density $k(x, t+\Delta t)$ and the calculated density $\widehat{k}$ using Eq.\ref{eq:CTM_single}. The fitness value for $N$ density quartets is expressed as:
\begin{equation}
    RMSE = -\sqrt{\frac{1}{N}\sum_{i=0}^{N}\left (k^i(x, t+\Delta t) - \widehat{k^i}  \right )^2}
\label{eq:rmse_fitness_FDestim}
\end{equation}

\subsection{Density Estimation} \label{sec:Density_Estimation}
To propagate traffic states on a link using the CTM, it is crucial to establish boundary condition values. The boundary conditions include: initial cell densities $k(x, t=0) \forall x$; inflow into the link: $Q(x=0,t) \forall t$; and outflow from the link: $Q(x=L,t) \forall t$. Using these values and the FD parameters, the traffic states on the link at each time-step can be estimated by propagating the initial cell densities.

\begin{figure}[htbp]
\centering
{\resizebox*{13cm}{!}{\includegraphics{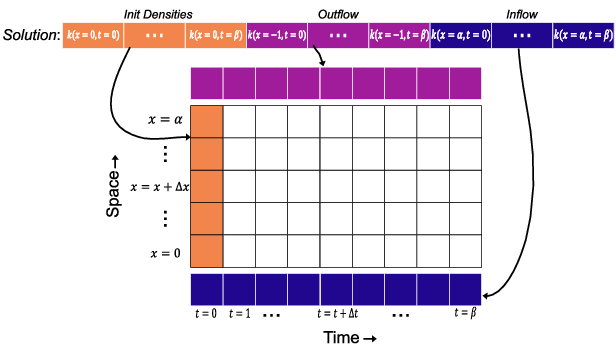}}}
\caption{Illustration showing the arrangement of a vector solution used to simulate CTM in the GA estimation algorithm.}
\label{fig:GA_candidate}
\end{figure}

In this step of the proposed methodology, we estimate the boundary conditions for each space-time density matrix using the GA method described in Section \ref{sec:GA}. The objective is to find a combination of boundary conditions that can accurately estimate the observed cell densities using CTM. In this particular GA implementation, each possible solution is represented as a vector consisting of initial cell densities, inflow, and outflow. The elements in the vector are sampled from a uniform distribution between 0 and $k_j$ (jam density). The inflow and outflow values are also sampled as densities, and the flows are calculated using a triangular FD function $Q(k(x,t),t)$ during the CTM process. Figure \ref{fig:GA_candidate} illustrates the arrangement of the solution vector used in the GA.

\newpage
To calculate the fitness of each solution, we execute the CTM with the calibrated FD parameters and boundary conditions to generate a density matrix $\mathbf{\widehat{k}^i}$ of dimension $(\alpha \times \beta)$, where $\alpha$ and $\beta$ represent the discretized number of space and time cells. This resultant density matrix is then compared with the actual density matrix $\mathbf{k}$ containing observed density cells. The fitness value for the solution is determined as the negative root mean square error (RMSE) calculated between the actual and CTM-calculated matrices. The fitness value for the $i^{th}$ solution is expressed as:

\begin{equation}
    RMSE = -\sqrt{\frac{1}{(\alpha\beta)}\sum_{x=0}^{\alpha}\sum_{t=0}^{\beta}(\mathbf{k} - \mathbf{\widehat{k}^i})^2}
\label{eq:rmse_fitness}
\end{equation}

This approach enables us to determine the optimal boundary conditions that minimize the discrepancy between observed and estimated traffic states. After calibrating the FD values and estimating the boundary conditions, we simulate the CTM with these optimized parameters. This final step generates a complete space-time diagram, estimating the previously unobserved cells.

\section{Simulation based Experiments} \label{sec:experiments}
To establish the reliability and accuracy of the proposed method, we require comprehensive information about traffic states on the link. Our approach is designed to leverage the unique data stream from on-board vehicle cameras, a concept detailed in Section \ref{sec:kitti}. However, validating this method in a real-world environment presents a significant data collection challenge. A proper evaluation would necessitate a specialized dataset featuring multiple runs of a camera-equipped vehicle along the same road segment, perfectly synchronized with ground-truth traffic state measurements (e.g., from loop detectors or other sensors) for the exact same link and timestamps. To our knowledge, no such public dataset exists, and its creation falls outside the scope of this research.

Due to the unavailability of suitable real-world data for validation, we utilize data generated through SUMO (Simulation of Urban MObility) (\cite{Lopez2018}), an open-source microscopic road traffic simulator. This simulation-based approach allows us to model a broad range of traffic scenarios and conditions that are difficult to capture through real-world data collection, thereby enabling a comprehensive evaluation of the proposed TSE method. By simulating the network under real-world conditions, we can generate various traffic scenarios for both incoming and parallel lane traffic, providing the flexibility needed for a thorough and robust assessment of our TSE method across a wide spectrum of traffic conditions.

\subsection{Simulation Setup}\label{sec:setup}
To validate our proposed methodology, we utilize the multi-modal traffic simulation of Ingolstadt, which is freely available in SUMO. The modal split encompasses passenger vehicles, public transport, bicycles, and pedestrians. This microscopic traffic simulation models traffic conditions across the entire network for a 24-hour period on September 16, 2020. The traffic flow for the "Ingolstadt model" is calibrated using real-world observations, with all traffic demand based on origin-destination matrices derived from mobility data from the city of Ingolstadt and Germany (\cite{Harth2022}).

\begin{table}[htbp]
\centering
\caption{Information about the three links used for validation.}
\label{tab:links}
\scriptsize
\begin{tabularx}{\textwidth}{>{\raggedright\arraybackslash}p{2cm} *{3}{>{\centering\arraybackslash}X}}
\toprule
\rule{0pt}{1cm} 
  & \includegraphics[width=0.9\linewidth,height=3.5cm]{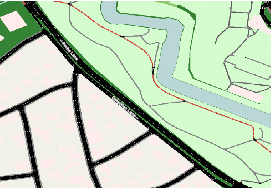} 
  & \includegraphics[width=0.9\linewidth,height=3.5cm]{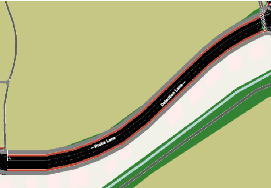} 
  & \includegraphics[width=0.9\linewidth,height=3.5cm]{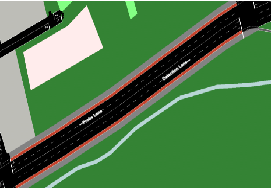} \\

\textbf{Attribute}        & \textbf{Link 1}     & \textbf{Link 2}     & \textbf{Link 3}     \\
\midrule
Street Name              & Westliche Ringstraße & Geroflinger Straße & Schloßlande         \\
Length [m]               & 430                  & 312                & 215                 \\
Direction                & S to N               & W to E             & NE to SW            \\
Lanes (Probe)            & 2                    & 1                  & 2                   \\
Lanes (Det.)             & 1                    & 2                  & 2                   \\
\midrule
\multicolumn{4}{c}{\textbf{Fundamental Diagram}} \\
\midrule
$v_f$ [km/h]             & 42                   & 44                 & 42                  \\
$k_c$ [veh/km]           & 44                   & 41                 & 60                  \\
$Q_{\text{max}}$ [veh/h] & 1848                 & 1804               & 2520                \\
\midrule
\multicolumn{4}{c}{\textbf{Experiment Parameters}} \\
\midrule
Probe runs               & 1477                 & 362                & 292                 \\
\bottomrule
\end{tabularx}
\end{table}

\begin{figure}[htbp]
\centering
{\resizebox*{11cm}{!}{\includegraphics{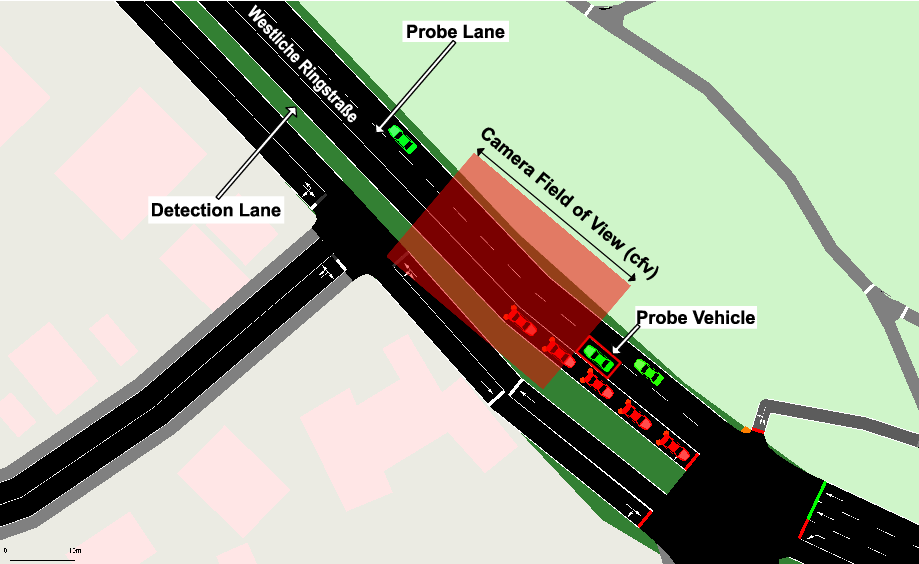}}}
\caption{Illustration of the setup from the Ingolstadt network used for data collection and validation. The RED shaded box represents the camera's field of view for probe vehicles. GREEN coloured vehicles are the camera-equipped probe vehicles on the Probe Lane, and they detect RED coloured vehicles on the Detection Lane.}
\label{fig:link}
\end{figure}

We selected three links from the Ingolstadt road network from which trajectory data is collected to validate the proposed methodology. Table \ref{tab:links} presents the details of selected links from the road network. We also present the flow density plots for the links in Figure \ref{fig:FD-Link}, alongside the triangular FD, calibrated from the SUMO simulation data. We replicate the camera-based trajectory data presented in Section \ref{sec:kitti} using the SUMO simulation. Figure \ref{fig:link} illustrates the setup used to collect trajectory data. In the simulation, we assume that camera-equipped probe vehicles, depicted in GREEN, are operating on the lane marked "Probe." Only a fraction of vehicles on this lane are equipped with cameras, controlled by a parameter called the camera penetration rate, which is fixed at 10\% for all links. The probe vehicles detect vehicles, depicted in RED, on the opposite lane marked as "Detection." This is the lane for which we capture the trajectories of vehicles within the camera's field of view ($cfv$), represented by a shaded red box ahead of the probe vehicle and is assumed to be 140 meters ahead of the probe vehicle. Vehicles on the detection lane, within this shaded red box, are considered for trajectory extraction.

\begin{figure}[ht]
\centering
\subfloat[Ground truth trajectories]{\resizebox*{7cm}{!}{\includegraphics{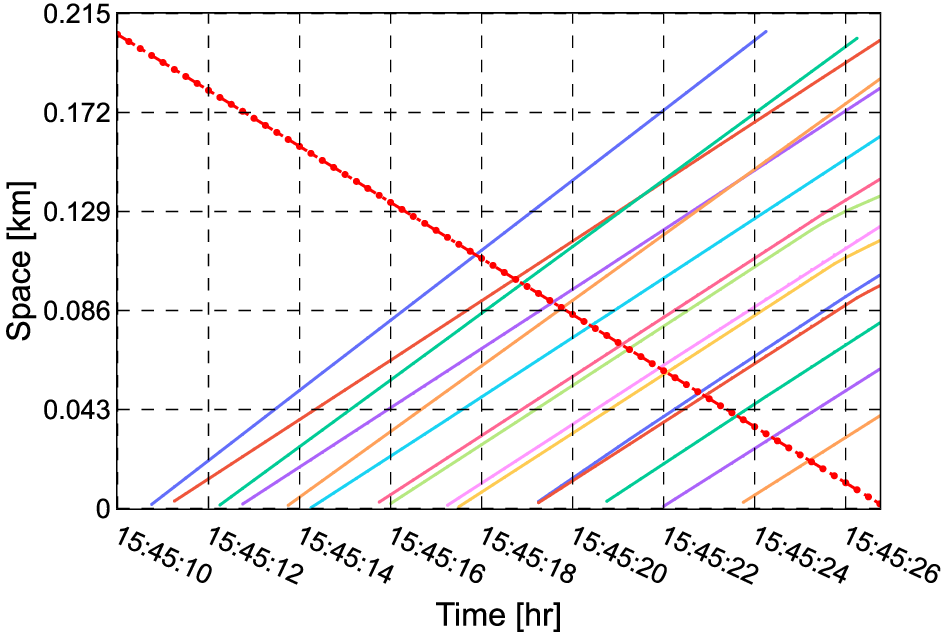}}\label{fig:GND_Contineous}}\hspace{1pt}
\subfloat[Camera measured trajectories]{ \resizebox*{7cm}{!}{\includegraphics{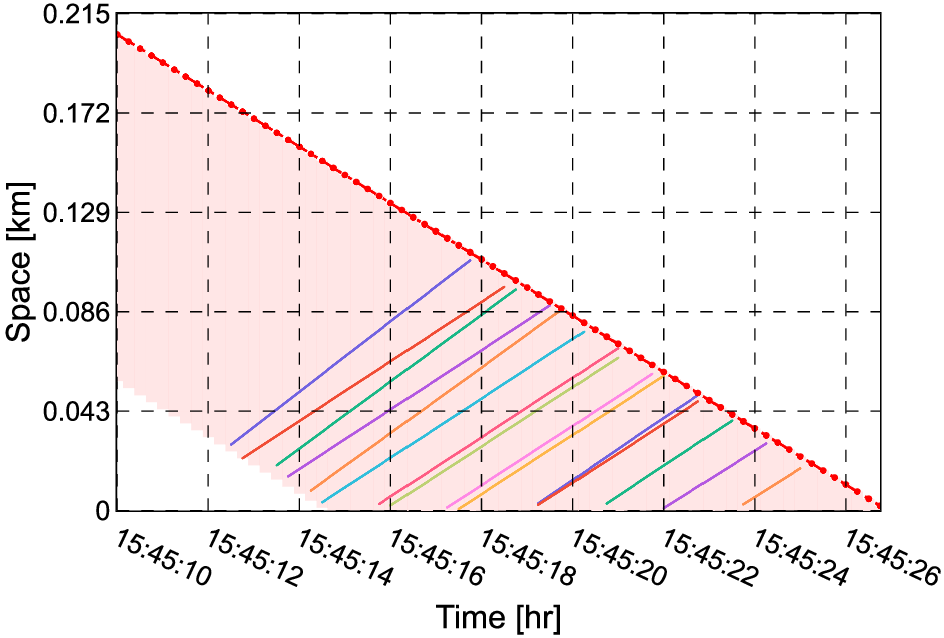}}\label{fig:Partial_Contineous}}\hspace{1pt}
\subfloat[Ground truth matrix]{ \resizebox*{7cm}{!}{\includegraphics{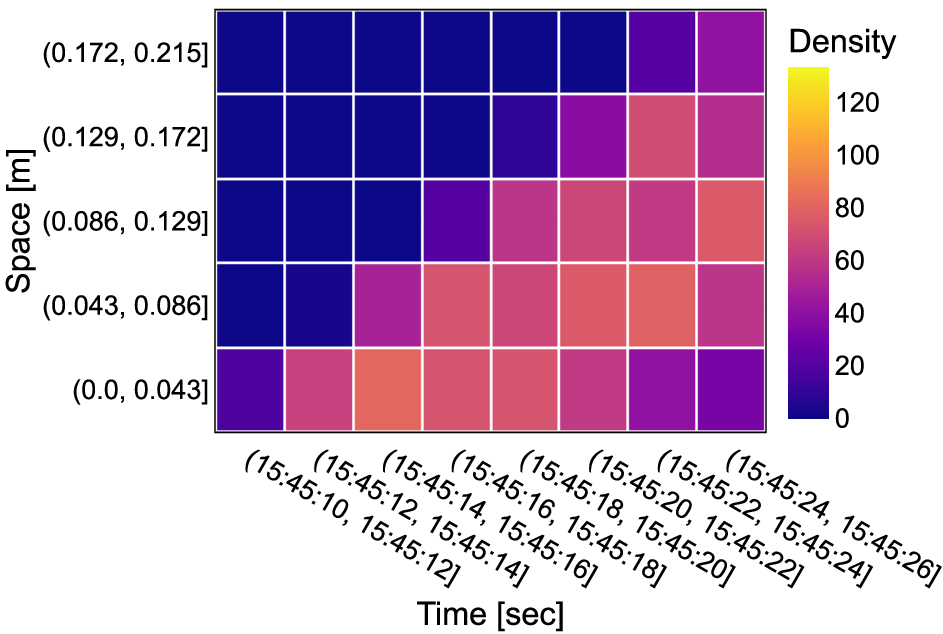}}\label{fig:GND_Matrix}}\hspace{1pt}
\subfloat[Partial matrix]{\resizebox*{7cm}{!}{\includegraphics{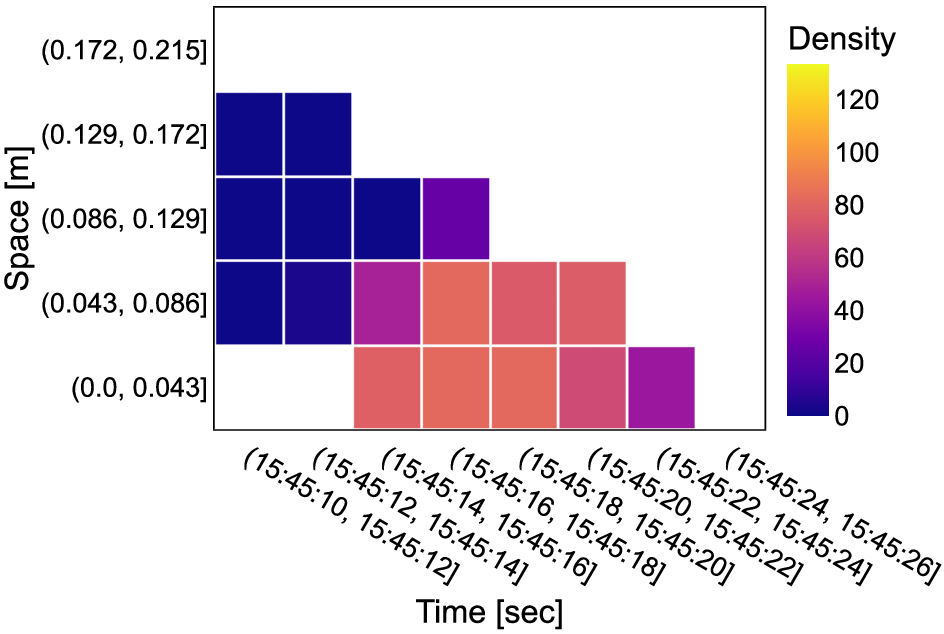}}\label{fig:Partial_Matrix}}\hspace{1pt}
\caption{Example of probe run derived from the SUMO simulation to showcase the proposed methodology. The trajectories are aggregated with $\Delta x = 43 m$ and $\Delta t = 2 sec$.}
\label{fig:example_sumo}
\end{figure}

Based on the penetration rate for each link, a specific number of unique probe runs are generated, each producing a distinct traffic space-time diagram for the detection link. The trajectory data from these probe runs are then discretized into a space-time density matrix using the method outlined in Section \ref{sec:aggregation}. For all experiments, the camera coverage per cell ($\delta_i$) is set at 50\%. Figure \ref{fig:example_sumo} presents an example space-time diagram, showing both ground-truth data and simulated camera-captured data with their corresponding discretized density matrices. In the trajectory plots, the bold red line represents the path followed by the probe vehicle on the probe link, while other lines depict the trajectories of vehicles in the detection link. The trajectories within the red-shaded region in the plot correspond to the $cfv$, while the partial trajectories represent the visible trajectories that will be utilized in the proposed TSE methodology.

\newpage
\subsection{Experimental Result}
This section focuses on evaluating the proposed TSE method using SUMO-simulated data for all three links. Specifically, we aim to assess the method's ability to generate Triangular FD parameters and boundary conditions for each space-time diagram using only partial matrices as input data. For the remainder of the analysis, the values presented in Table \ref{tab:delta-values} for $\delta x$ and $\delta t$ are used to discretize the respective space-time diagrams. Additionally, we report the jam density for each link, where $k_j$ is determined as the inverse of the minimum headway distance, set to 7.5 meters in the SUMO simulation. We begin by presenting the results of the FD calibration step, and subsequently apply the calibrated FD to the density estimation step in order to estimate unobserved density cells.

\begin{table}[ht]
\centering
\caption{Discretization values and jam density for links in experiment.}
\label{tab:delta-values}
\begin{tabularx}{\textwidth}{@{}lXXX@{}}
\toprule
           & \textbf{Link 1} & \textbf{Link 2} & \textbf{Link 3} \\ \midrule
$\Delta x$ & 0.086 km        & 0.052 km        & 0.043 km        \\
$\Delta t$ & 3 sec           & 3 sec           & 2 sec           \\
$k_j$      & 133.33 veh/km   & 266.67 veh/km   & 266.67 veh/km   \\ \bottomrule
\end{tabularx}
\end{table}

\subsubsection{Result: FD Calibration}
The FD calibration step uses density quartets, which are explained in detail in Section \ref{sec:FD_estimation}, as input for the proposed GA. Using the partial density matrices from unique probe runs, we were able to extract a number of quartets. We generated a paired sample of $v_f$ and $k_c$ from a uniform distribution, subject to constraints given by Eq.\ref{eq:cfl} and Eq.\ref{eq:optimDensityConstrain}.

The GA utilized in our FD calibration process requires meticulous parameter tuning to achieve optimal performance. The set of GA hyper-parameters used for each link was established through extensive search and is provided in Table \ref{tab:FD-Result}. It's important to highlight that although these parameters were effective for our specific case, the GA demonstrated a degree of robustness to minor variations in their values. We selected the best output from these runs as our final result.

The GA-estimated FD values for each link are presented in Table \ref{tab:FD-Result}, along with the GA hyper-parameters and the absolute difference from the actual values. These estimated values closely align with the known values for both $v_f$ and $k_c$, as evidenced by the visual comparison in Figure \ref{fig:FD-Link}. The plots demonstrate a strong correlation between the estimated and known FD values. The outcomes of the FD calibration phase suggest that our proposed GA-based method can accurately and efficiently estimate Triangular FD parameters using limited data from space-time diagrams. This successful calibration lays a robust groundwork for the subsequent density estimation steps.

\subsubsection{Result: Density estimation}
Using the calibrated Triangular FD parameters from the previous step, we now focus on estimating the boundary conditions for each space-time matrix. By identifying these boundary conditions, we can provide the necessary inputs for the CTM to estimate the unobserved density cells throughout the entire space-time diagram.

As shown in Section \ref{sec:Density_Estimation}, we start the proposed GA process by making an initial population. Each solution in this population is a vector combination of boundary conditions. The vector values are randomly selected from a uniform distribution within the range $[0, k_j]$, where $k_j$ represents the jam density. Through extensive experimentation, we established the set of GA parameters for each link. Given that the GA procedure does not guarantee a globally optimal solution, we conducted multiple independent runs with the same parameters to ensure consistent outcomes. The best output from these runs was selected as the final result, enhancing the reliability of our estimation process. Once the optimal boundary conditions are obtained from the GA, they are combined with the calibrated FD values to generate the density cell values for the entire space-time domain. This represents the final outcome of the proposed methodology. This process is repeated for each of the space-time matrices derived from all the probe runs in the experiment.

To evaluate the model's performance, we calculate the RMSE between the ground truth matrices and the GA output matrices for each scenario, specifically focusing on the cells located below the trajectory of the probe vehicle. The reported RMSE values encompass all cell densities, including initial densities as well as inflow and outflow values that have been converted into density cells. Our analysis is restricted to the cells beneath the probe vehicle's trajectory due to the lack of information regarding outflow in the opposite lane, which makes it impossible to predict traffic flow beyond the passage of vehicles in that direction. As a result, cells outside this region are excluded from consideration. Appendix \ref{appendix:matrix-shape} provides an example scenario in which the GA-estimated space-time diagram is compared to the ground truth. Table \ref{tab:Density-Result} presents the GA hyper-parameters along with the mean-RMSE calculate for all probe runs. As a visual inspection, we present the two distributions of RMSE in Figure \ref{fig:RMSE-Link}: 1.) error between known matrices and GA-estimated matrices. 2.) error between know matrices and CTM-estimated matrices. The CTM-estimated matrices are generated using the known FD and boundary conditions and serve as the basis for comparison.

The low mean-RMSE values across all probe runs for the three links demonstrate that our proposed density estimation phase can generate reliable boundary conditions for estimating unobserved density cells. This reliability is further corroborated by the close similarity between the RMSE distributions of GA-estimated boundary conditions and known boundary conditions. These findings provide strong evidence for the effectiveness of our approach.

\newpage
\begin{landscape}
    \begin{minipage}[t]{0.85\textheight}
        \begin{figure}[H]
            \centering
            \begin{tabular}{cc}
                \includegraphics[width=5.2cm]{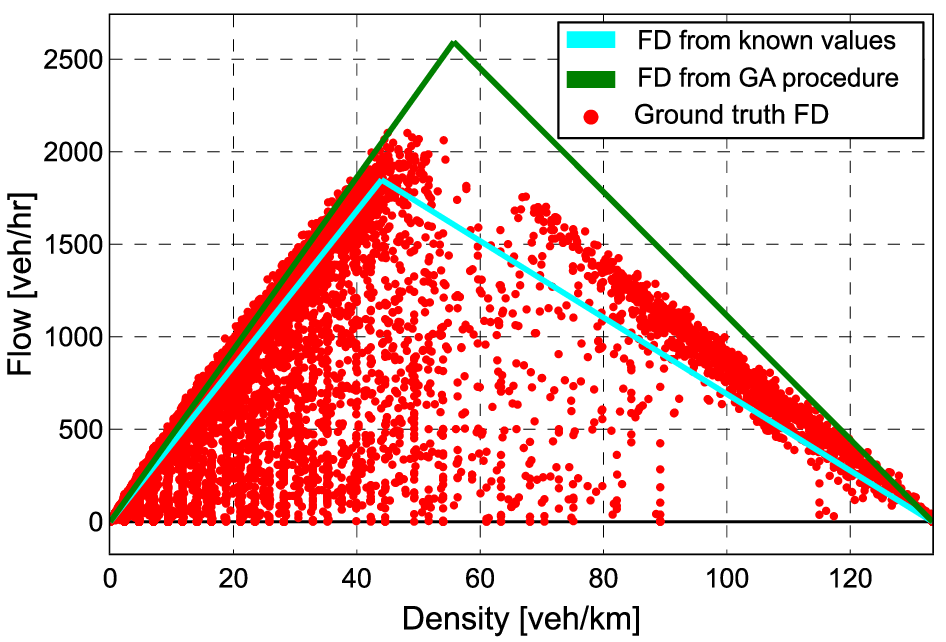} &
                \includegraphics[width=5.2cm]{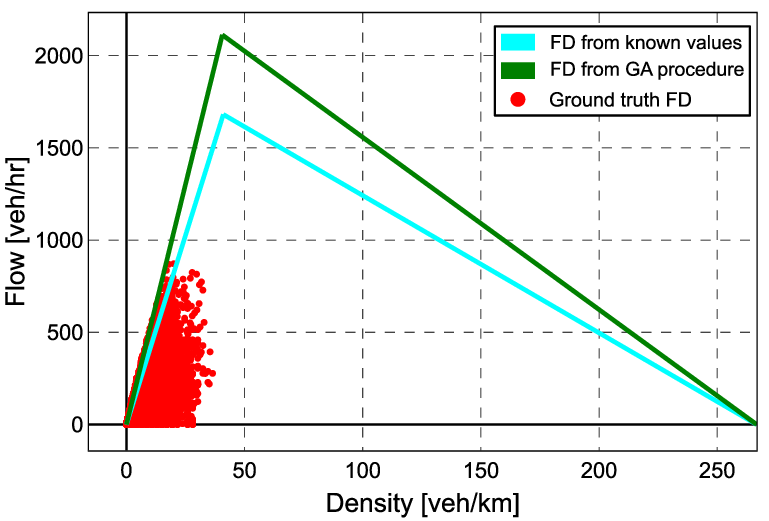} \\
                \multicolumn{1}{c}{\small (a) Link 1} & 
                \multicolumn{1}{c}{\small (b) Link 2} \\
            \end{tabular}
            
            \begin{tabular}{c}
                \includegraphics[width=5.2cm]{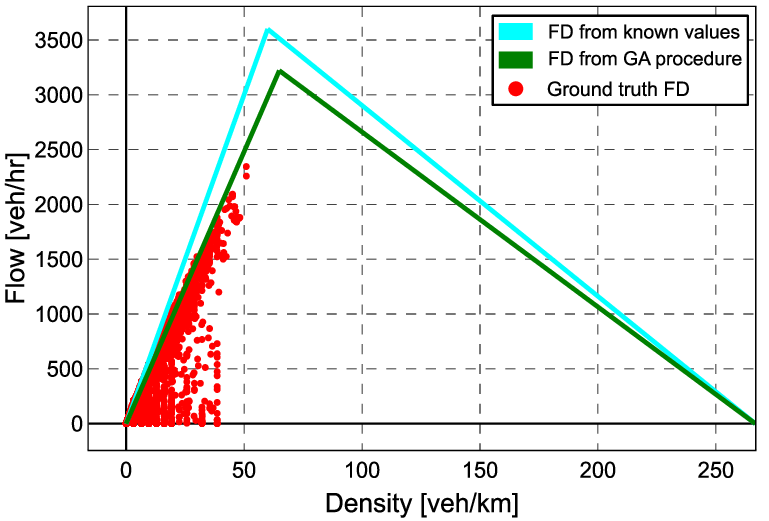} \\
                \small (c) Link 3 \\
            \end{tabular}
            
            \caption{Flow-density curve for each link showcasing the known triangular FD (in cyan) as well as the FD estimated using the proposed methodology (in green).The RED dot represents the FD derived directly from the SUMO simulation at 2 second resolution.}
            \label{fig:FD-Link}
        \end{figure}
    \end{minipage}
    \hfill
    \begin{minipage}[t]{0.6\textheight}
        \centering
        \captionof{table}{Result of the FD calibration on the 3-links.}
        \resizebox{\textwidth}{!}{%
            \begin{tabular}{@{}lccc@{}}
                \toprule
                & \textbf{Link 1} & \textbf{Link 2} & \textbf{Link 3} \\ \midrule
                N quartets & 4080 & 1896 & 1745 \\ \midrule
                \multicolumn{4}{c}{\textbf{GA Hyperparameters}} \\ \midrule
                N Generation & 100 & 100 & 100 \\
                K-tournament & 15 & 7 & 15 \\
                Adaptive mutual prob. & 0.1, 0.9 & 0.5, 0.5 & (0.25, 0.75) \\
                Crossover fraction & 5\% & 20\% & 10\% \\ \midrule
                \multicolumn{4}{c}{\textbf{FD Estimation Result}} \\ \midrule
                Estim. $v_f$ & 46.55 km/hr & 52.06 km/hr & 49.61 km/hr \\
                Estim. $k_c$ & 55.71 veh/km & 40.57 veh/km & 64.92 veh/km \\ \midrule
                \multicolumn{4}{c}{\textbf{Absolute Difference from Known Values}} \\ \midrule
                $\Delta v_f$ & 4,55 & 8,06 & 7,61 \\
                $\Delta k_c$ & 11,71 & 0,43 & 4,92 \\ \bottomrule
            \end{tabular}
        }
        \label{tab:FD-Result}
    \end{minipage}
\end{landscape}

\newpage
\begin{landscape}
    \begin{minipage}[t]{0.75\textheight}
        \begin{figure}[H]
            \centering
            \begin{tabular}{cc}
                \includegraphics[width=4cm]{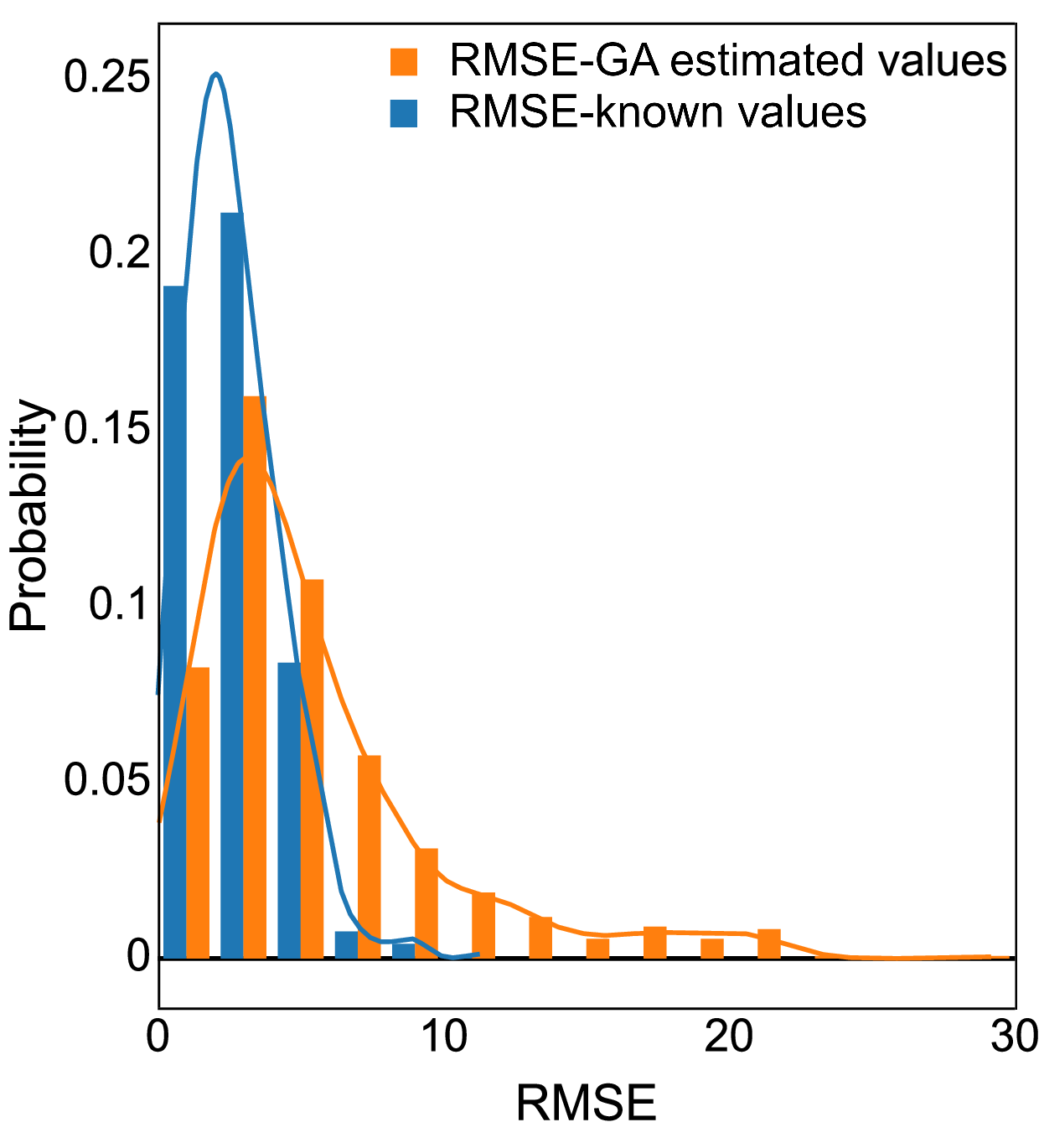} &
                \includegraphics[width=4cm]{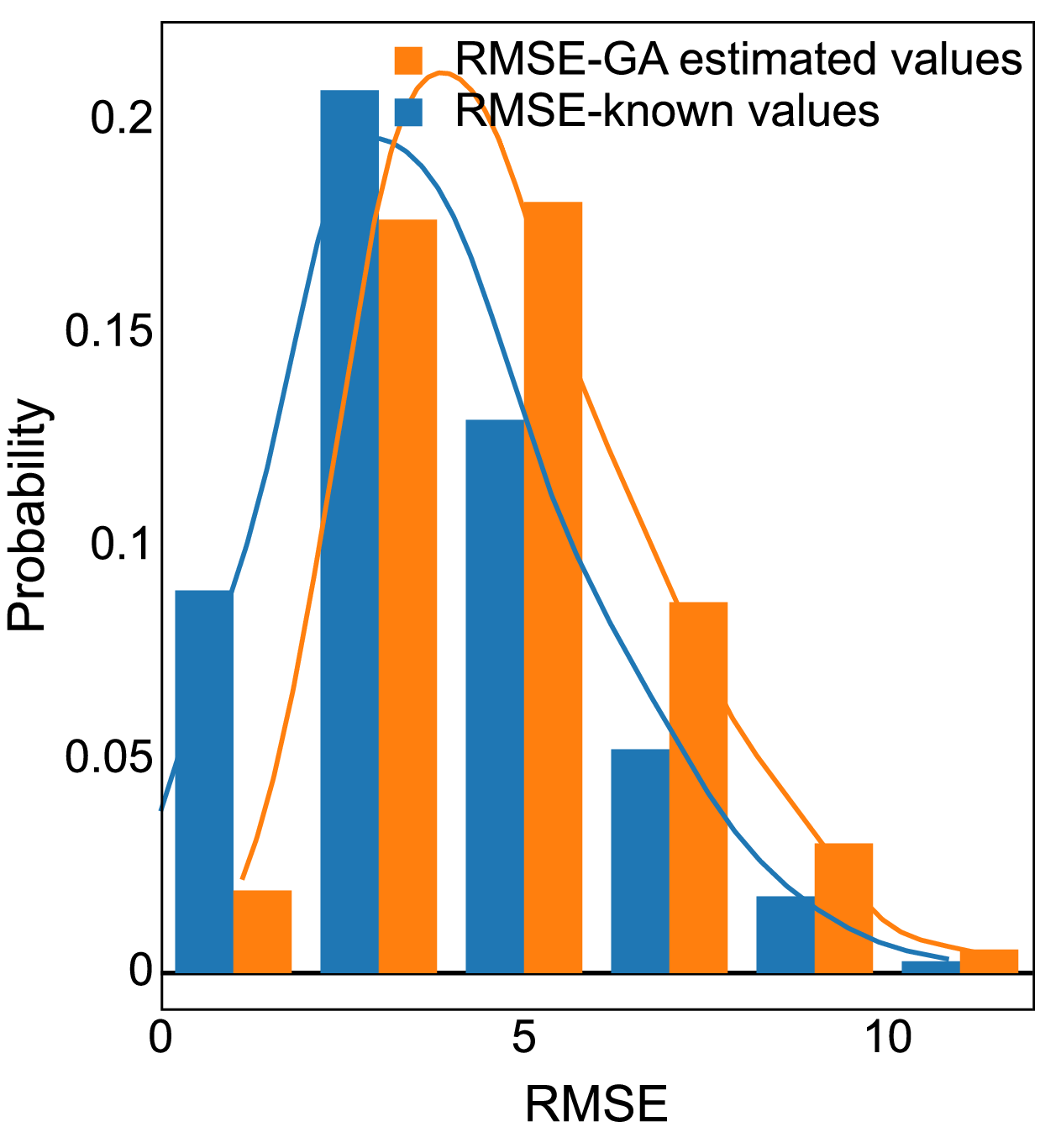} \\
                \multicolumn{1}{c}{\small (a) Link 1} & 
                \multicolumn{1}{c}{\small (b) Link 2} \\
            \end{tabular}
            
            \begin{tabular}{c}
                \includegraphics[width=4cm]{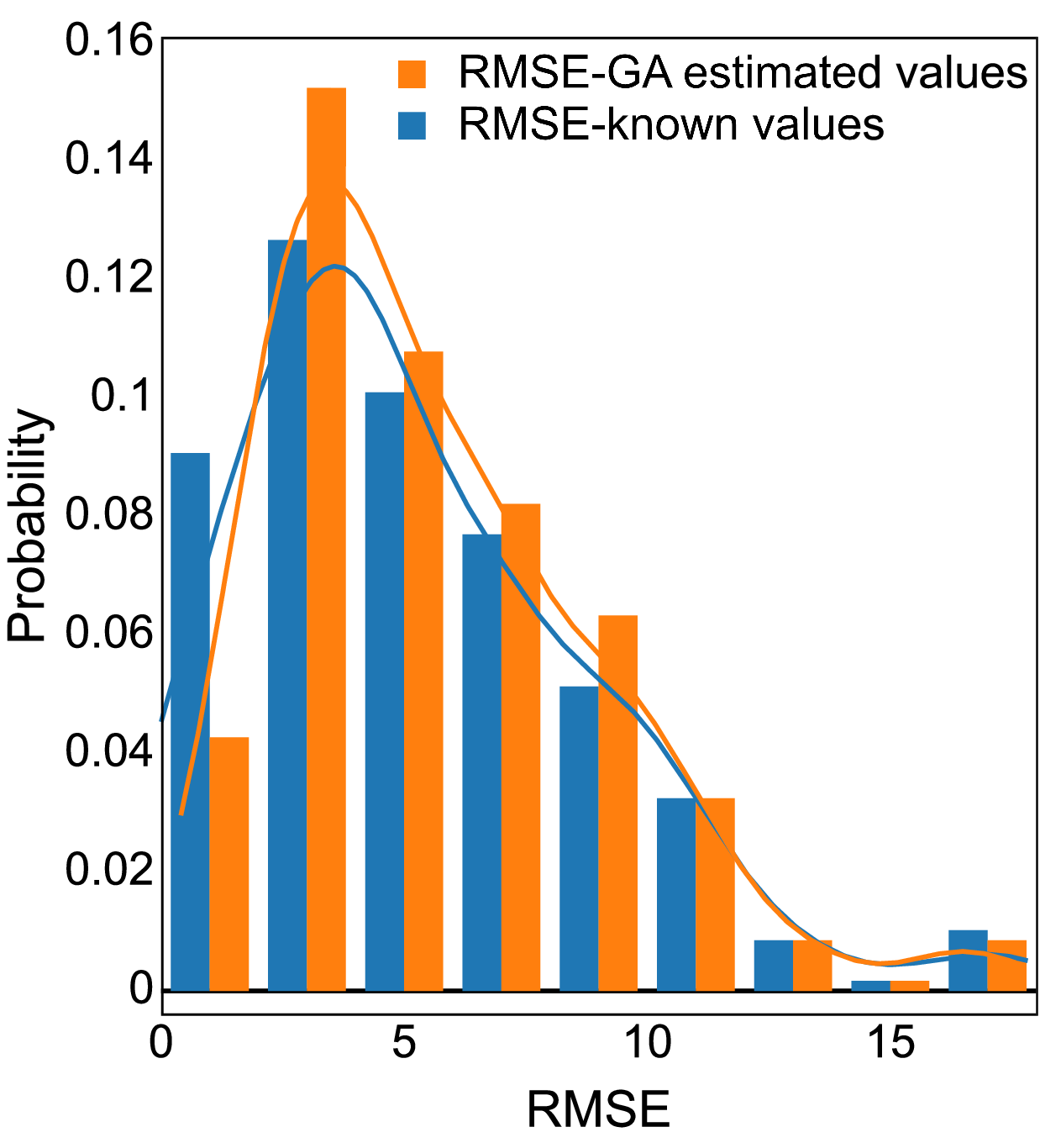} \\
                \small (c) Link 3 \\
            \end{tabular}
            
            \caption{The RMSE curve was computed by comparing the ground truth to the matrices with GA estimated matrices (in orange) and known value matrices (in blue).}
            \label{fig:RMSE-Link}
        \end{figure}
    \end{minipage}
    \hfill
    \begin{minipage}[t]{0.7\textheight}
        \centering
        \captionof{table}{Result of Density estimation on the 3-links.}
        \resizebox{\textwidth}{!}{%
            \begin{tabular}{@{}lccc@{}}
            \toprule
                                  & \textbf{Link 1}  & \textbf{Link 2}  & \textbf{Link 3}  \\ \midrule
            N probe runs          & 1477             & 362              & 292              \\ \midrule
            \multicolumn{4}{c}{\textbf{GA Hyperparameters}}                                \\ \midrule
            N Generation          & 100              & 100              & 100              \\
            Population Size       & 200              & 200              & 200              \\
            K-tournament          & 2                & 9                & 5                \\
            Adaptive mutual prob. & 0.1, 0.9         & 0.5, 0.5         & (0.25, 0.75)     \\
            Crossover fraction    & 50\%             & 50\%             & 50\%             \\ \midrule
            \multicolumn{4}{c}{\textbf{Density Estimation Error}}                          \\ \midrule
            mean-RMSE             & 5.28+3.84 veh/km & 4.82+1.93 veh/km & 5.62+3.31 veh/km \\ \bottomrule
            \end{tabular}
        }
        \label{tab:Density-Result}
    \end{minipage}
\end{landscape}

\subsection{Result Analysis}
The results for both FD calibration and density estimation phases indicate better performance on Link 2 and Link 3 compared to Link 1. As shown by the results, GA-estimated values align more closely with known values for Links 2 and 3. This is likely due to the smaller dataset and limited variability in the simulated data available for Links 2 and 3, where data is only captured for the free-flow segment of the true FD. In contrast, Link 1 includes data spanning the entire true FD.

To understand these performance differences, the following sections provide a detailed analysis of potential error sources. This analysis examines the data collection process, the algorithmic design, and the underlying modeling assumptions that may contribute to the observed discrepancies. Our investigation is divided into two primary areas. First, Sections \ref{sec:data_pen_rate} and \ref{sec:data_camera} discuss how the characteristics and limitations of the data collection process—specifically, its reliance on trajectory data from vehicles in the opposing lane—influence the method's effectiveness. Second, Sections \ref{sec:ctm_limitations} and \ref{sec:ctm_GA} explore how the performance is governed by the chosen modeling and estimation framework, which combines the CTM with GA.

Before delving into this analysis, it is important to acknowledge that our evaluation relies on SUMO-simulated data. Although these simulations were calibrated with real-world data, they may not fully capture the complex dynamics of actual traffic conditions, as illustrated by the real-world examples in Section \ref{sec:kitti}. This simplification can lead to disparities between simulated outcomes and on-street traffic states, potentially affecting the practical applicability of our findings. Acknowledging this limitation provides crucial context for the following discussion on error sources and their implications.

\subsubsection{Practical Considerations}
The proposed estimation method operates offline, processing data only after sufficient probe runs have been collected for a given link. Although individual computational steps are fast, the overall runtime increases due to repeated application across thousands of probe runs and hundreds of GA generations (see Appendix \ref{appendix:runtimes}). However, this computational burden can be substantially reduced by leveraging the parallelizable nature of GA, which is well-suited to multi-threaded or multi-processor environments. Additionally, once near-optimal solutions are found, future estimations require far fewer iterations, significantly lowering subsequent computational costs.

\subsubsection{Effect of Penetration Rate} \label{sec:data_pen_rate}
In the FD calibration phase, the proposed GA method demonstrated its capability to accurately calibrate the Triangular FD parameters using a limited number of density quartets. The estimated values for $v_f$ and $k_c$ were well within a reasonable range of error from the actual FD parameters across all three links used in the validation. However, the accuracy of the FD calibration step heavily depends on incorporating a large number of diverse quartets. The extraction of these quartets from the density matrices is influenced by two key factors: the penetration rate of camera-equipped probe vehicles and the discretization of space-time diagrams ($\Delta x$). Higher penetration rates and smaller $\Delta x$ values lead to a greater number of quartets. Crucially, these quartets should represent traffic states from different parts of the true link FD to ensure comprehensive calibration.

Table \ref{tab:prate_deltaX} shows the total number of quartets that were extracted along with their mean-RMSE values when the penetration rate and $\Delta x$ values were varied from known density matrices from all three probe runs. We calculated the RMSE values by simulating the CTM with the FD parameters using the method described in Section \ref{sec:FD_estimation}. The mean-RMSE is then computed across all quartet values for each combination of penetration rate and $\Delta x$ value, providing insight into the impact of these factors on calibration accuracy.

\begin{table}[htbp]
\centering
\caption{The mean-RMSE value, calculated for all probe runs at different combinations of penetration rate and cell size. The numbers in the bracket represent the number of quartets extracted for that combination.}
\label{tab:prate_deltaX}
\begin{tabularx}{\textwidth}{@{}c*{5}{>{\centering\arraybackslash}X}@{}}
\toprule
\multirow{2}{*}{\textbf{\begin{tabular}[c]{@{}c@{}}Penetration\\ Rate\end{tabular}}} &
  \multicolumn{2}{c}{\textbf{Link 1}} &
  \multicolumn{2}{c}{\textbf{Link 2}} &
  \textbf{Link 3} \\ \cmidrule(l){2-6} 
 &
  dX - 0,043 &
  dX - 0,086 &
  dX - 0,039 &
  dX - 0,052 &
  dX - 0,043 \\ \midrule
0.1 &
  \begin{tabular}[c]{@{}c@{}}4.97\\ (227 116)\end{tabular} &
  \begin{tabular}[c]{@{}c@{}}6.73\\ (165 175)\end{tabular} &
  \begin{tabular}[c]{@{}c@{}}2.37\\ (29 809)\end{tabular} &
  \begin{tabular}[c]{@{}c@{}}3.54\\ (17 885)\end{tabular} &
  \begin{tabular}[c]{@{}c@{}}9.31\\ (6 725)\end{tabular} \\
0.15 &
  \begin{tabular}[c]{@{}c@{}}4.13\\ (576 168)\end{tabular} &
  \begin{tabular}[c]{@{}c@{}}6.30\\ (480 972)\end{tabular} &
  \begin{tabular}[c]{@{}c@{}}2.29\\ (74 559)\end{tabular} &
  \begin{tabular}[c]{@{}c@{}}3.51\\ (56 683)\end{tabular} &
  \begin{tabular}[c]{@{}c@{}}4.65\\ (10 967)\end{tabular} \\
0.2 &
  \begin{tabular}[c]{@{}c@{}}3.62\\ (1 040 148)\end{tabular} &
  \begin{tabular}[c]{@{}c@{}}5.71\\ (913 608)\end{tabular} &
  \begin{tabular}[c]{@{}c@{}}2.17\\ (134 229)\end{tabular} &
  \begin{tabular}[c]{@{}c@{}}3.51\\ (119 337)\end{tabular} &
  \begin{tabular}[c]{@{}c@{}}4.59\\ (13 157)\end{tabular} \\ \bottomrule
\end{tabularx}%
\end{table}

The results clearly show that as the penetration rate increases, the mean RMSE value decreases across all three links. This trend indicates that a higher rate yields more scenarios from which quartets can be extracted, thereby improving the estimation process. Moreover, the number of quartets is highly dependent on the discretization cell size; for Links 1 and 2, larger spatial cells significantly reduce the number of available quartets. Crucially, for the estimation to improve, the extracted quartets must represent diverse parts of the true FD. This is confirmed by the results, which show significantly better improvement for Link 1, with data from the entire FD, compared to Links 2 and 3, which only have data from the free-flow section.

\newpage
This analysis demonstrates that the estimation results depend heavily on both the quantity and diversity of the quartets. However, achieving this diversity is challenging in real-world scenarios, where we have limited control over the type of data collected by probe cameras. This limitation highlights the importance of careful consideration in data collection and processing methods to ensure a comprehensive representation of traffic states across the entire FD spectrum.

\subsubsection{Camera Detection Constrains}\label{sec:data_camera}
The length of the extracted trajectory primarily depends on two factors: the detection capability of the object detector and the visibility of the detected vehicle on the camera. While the use of the state-of-the-art YOLOv5 object detector addresses the first factor, the limited distance at which a camera can detect vehicles ($cfv$) remains a constraint of this proposed method. This also restricts the value of $\Delta x$, as it cannot exceed the maximum $cfv$. For values of $\Delta x$ greater than the camera detection distance, none of the detected trajectories cover even a single space-time cell. Consequently, we cannot capture any tuples required for the FD-calibration step, precluding subsequent density estimation.

The constraint on $\Delta x$ also limits the permissible values of $v_f$ due to the CFL condition required for the proposed methodology. If the maximum value that $\Delta x$ can take is 150 m, then the $v_f$ values are constrained by: $v_f \cdot \Delta t \leq 0.15$. To satisfy this equation, either $v_f$ needs to be tiny for normal values of $\Delta t$, or $\Delta t$ must be reduced significantly. This restricts the applicability of the proposed method in locations where the free-flow speed is relatively high.

\subsubsection{Limitations of CTM Due to Discretization and Traffic Conditions} \label{sec:ctm_limitations}
The proposed methodology relies on CTM to estimate unobserved traffic densities within space-time matrices. While CTM offers simplicity and computational efficiency, it introduces notable limitations related to discretization and the assumptions it imposes on traffic flow homogeneity. Consequently, errors generated by the CTM propagate
through the proposed GA-based methodology.

CTM performance is highly sensitive to the choice of spatial ($\Delta x$) and temporal ($\Delta t$) discretization values used for continuous space-time diagrams. To analyze this sensitivity towards these parameters, we conducted simulations using known FD parameters and boundary conditions across various combinations of $\Delta x$ and $\Delta t$. For each configuration, we computed density matrices using CTM and compared them to ground-truth matrices derived from complete vehicle trajectories using Edie’s generalization. We employed RMSE as the evaluation metric to compare cell densities between the ground truth and simulated CTM matrices. Table \ref{tab:dtdx-link} reports the mean RMSE values for each link. We ensured all chosen discretizations satisfy the CFL condition, excluding invalid combinations. The results in the table indicate that the variation in CTM accuracy can largely be attributed to discretization. The results reveal that changes in $\Delta x$ generally have a more pronounced effect on estimation accuracy than $\Delta t$, as evident from the values for Links 1 and 2. These findings emphasize the influence of spatial resolution on the CTM's ability to replicate true traffic dynamics.

\begin{table}[htbp]
\centering
\caption{mean-RMSE for density matrices calculated for all links by varying the values of $\Delta x$ and $\Delta t$. The lowest value is highlighted in RED.}
\label{tab:dtdx-link}
\begin{tabularx}{\textwidth}{@{}c*{5}{>{\centering\arraybackslash}X}@{}}
\toprule
  & \multicolumn{2}{c}{\textbf{Link 1}}    & \multicolumn{2}{c}{\textbf{Link 2}}  & \textbf{Link 3}              \\ \midrule
\textbf{dT} & \textbf{dX - 0.043} & \textbf{dX - 0.086} & \textbf{dX - 0.039} & \textbf{dX - 0.052} & \textbf{dX - 0.043} \\ \midrule
2 & 11.950 & 10.686                        & 6.423 & 6.026                        & {\color[HTML]{FE0000} 7.580} \\
3 & 11.222 & {\color[HTML]{FE0000} 10.076} & 6.134 & {\color[HTML]{FE0000} 5.517} & 7.728                        \\
4 & -      & 10.198                        & -     & 5.698                        & -                            \\
5 & -      & 10.500                        & -     & -                            & -                            \\
\bottomrule
\end{tabularx}
\end{table}

These discrepancies stem from inherent limitations in Edie's generalization, which uses samples that may not map to the same fundamental diagram. For instance, expanding the area ($\mathbf{A_i}$) in Eq \ref{eq:density} to include vehicle-free sections increases the denominator while the numerator remains constant, artificially lowering the calculated density and flow. Furthermore, the aggregation of space-time diagrams into density cells assumes uniformity, overlooking transient changes and causing mismatches between CTM-generated values and actual data, particularly in capturing transient states.

\begin{figure}[htbp]
\centering
\subfloat[Link 1]{\resizebox*{6cm}{!}{\includegraphics{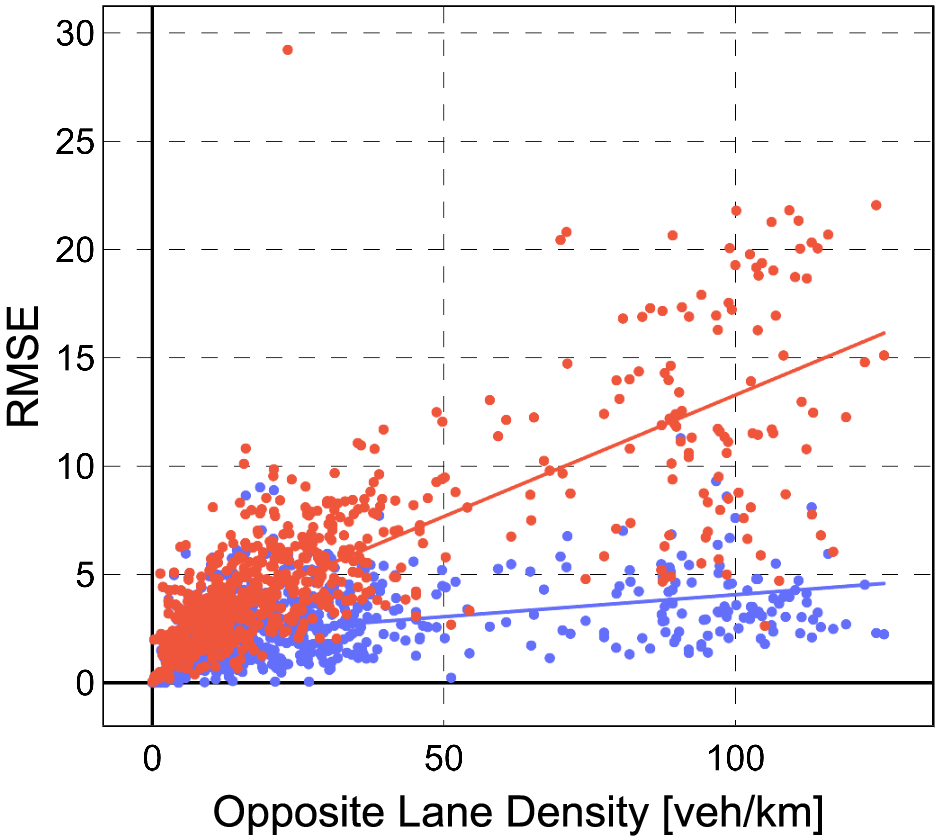}}\label{fig:LaneDensity-link1}}\hspace{1pt}
\subfloat[Link 2]{\resizebox*{6cm}{!}{\includegraphics{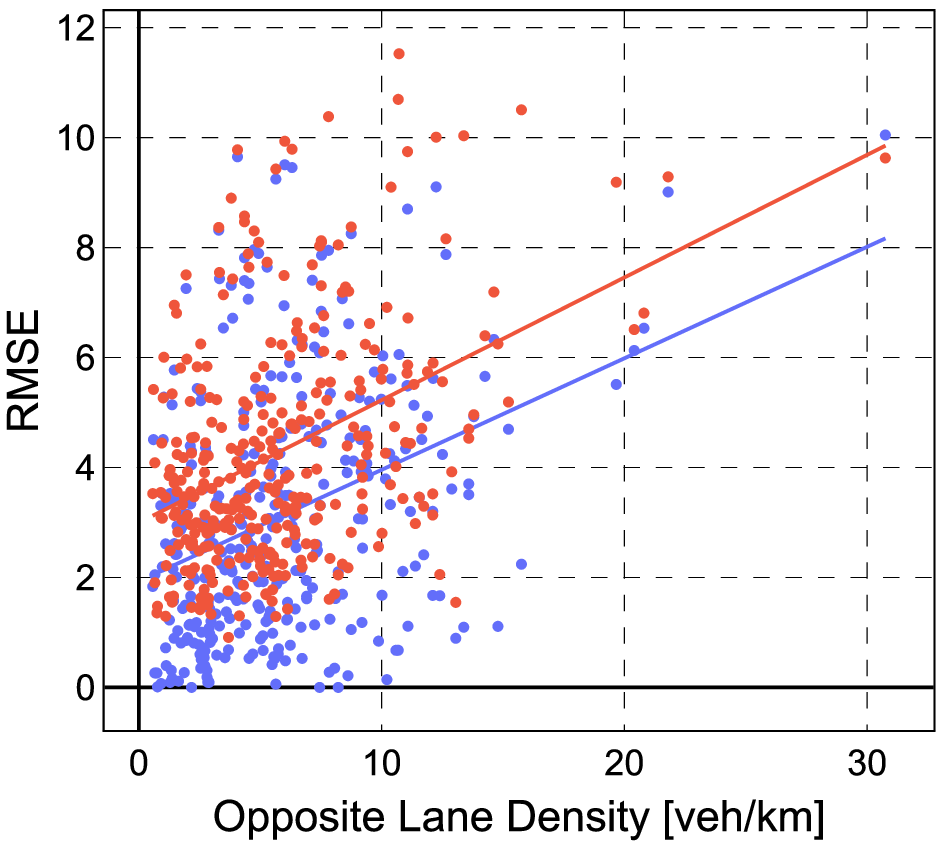}}\label{fig:LaneDensity-link2}}\hspace{1pt}
\subfloat[Link 3]{\resizebox*{6cm}{!}{\includegraphics{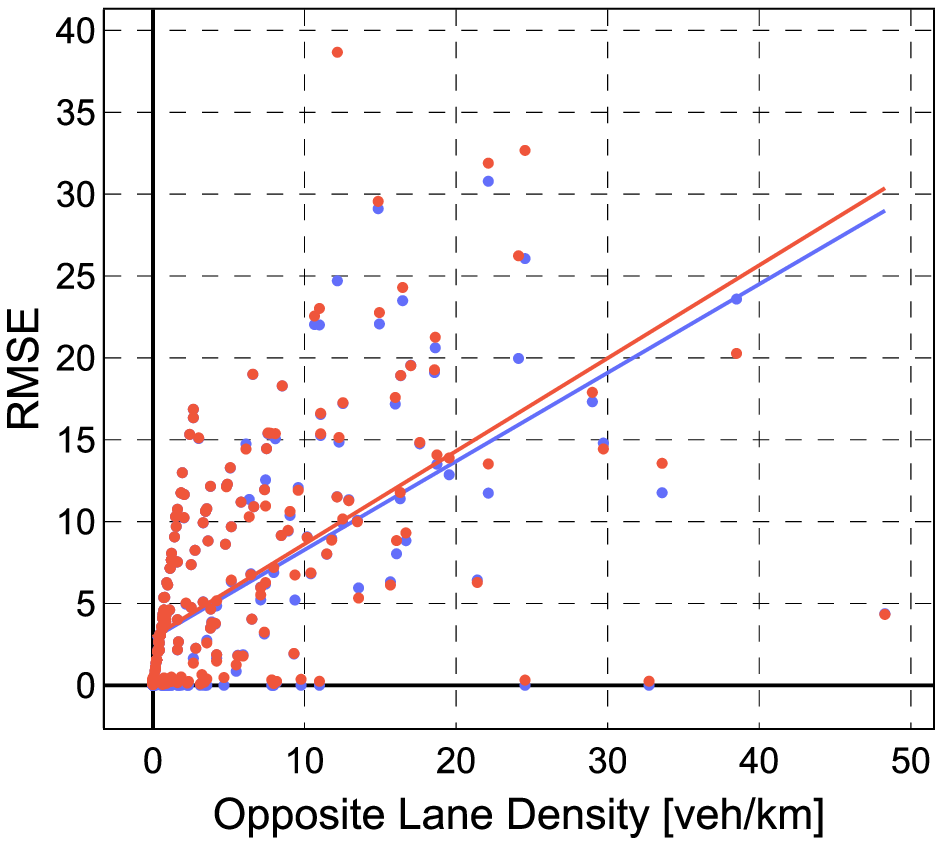}}\label{fig:LaneDensity-link3}}\hspace{1pt}
\caption{Plots showcasing density in the opposite lane versus the RMSE calculated, using known values (in BLUE) and GA estimated values (in RED) for all three links.}
\label{fig:RMSE_vs_trafficStates}
\end{figure}

The limitations of CTM are further magnified under varying traffic conditions. In the density estimation phase, we observed that RMSE values increased for probe runs exhibiting significant variations in lane density and flow. Figure \ref{fig:RMSE_vs_trafficStates} illustrates this trend, showing RMSE as a function of average link density across probe runs for all three links. RMSE values calculated using GA-estimated parameters (in red) are compared against those obtained using ground-truth parameters (in blue). 

A positive correlation between RMSE and average link density is evident, indicating that CTM struggles to capture transient states under higher traffic variability. This is especially noticeable in Link 1, where the data spans the full range of the FD, as opposed to Links 2 and 3, which cover only the free-flow regime. These challenges stem from CTM’s reliance on fixed cell boundaries and homogeneous state assumptions, which can distort representations of rapidly changing conditions.

To address these shortcomings, several more sophisticated traffic flow models could be considered. Potential alternatives include the Switching Mode Model (SMM) (\cite{Munoz2006}), the Stochastic CTM (\cite{Sumalee2011}), higher-order Lighthill-Whitham-Richards (LWR) models such as the one proposed by \cite{Newell1993}, or models that avoid state discretization altogether, like the Lax-Hopf-based method (\cite{Mazare2011}). A key advantage of our proposed methodology is its modular design, which allows the CTM to be simply replaced by one of these more advanced models. For instance, substituting the CTM with the SMM would likely yield more accurate results, given the SMM's excellent adaptability to anomalies and higher robustness to noise. However, this enhanced performance comes at the cost of greater implementation complexity, higher data requirements for reliable mode detection, and slower computational speed. In contrast, the CTM offers simplicity, high interpretability, and rapid computation. Therefore, the CTM was strategically selected for this initial study to prioritize implementation simplicity and computational efficiency, acknowledging its limitations while establishing a flexible framework for future enhancements.

\subsubsection{Density Estimation using GA} \label{sec:ctm_GA}
\begin{figure}[htbp]
\centering
{\resizebox*{13cm}{!}{\includegraphics{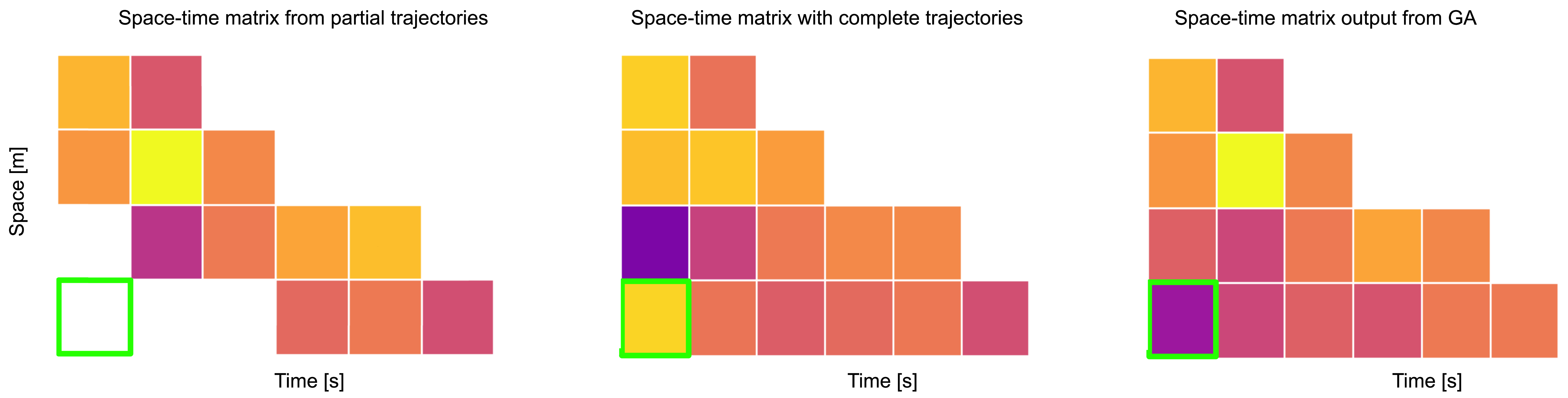}}}
\caption{This illustration illustrates how deviations from the initial density, highlighted by the "green" box, can lead to error propagation in other cells.}
\label{fig:init_cell_issue}
\end{figure}

When estimating the boundary values for CTM during the density estimation step, deviations from the actual initial densities can lead to an overall increase in RMSE due to error propagation. Fig. \ref{fig:init_cell_issue} provides an example where the green highlighted cell, generated by the GA, significantly deviates from the actual data. We observed that when the initial density cells are located further away from any observed density cells, the values generated by the GA are often suboptimal. As a result, when these suboptimal values are propagated through the CTM, incorrect cell values are generated in subsequent time steps, leading to an overall increase in the RMSE value.

\section{Conclusions}\label{sec:conclusion}
This paper tackles the significant challenge of estimating traffic states on road links using data from on-board vehicle cameras. The main insight from this study is the demonstrated potential of on-board cameras to deliver high-resolution traffic data across entire road links. We presented a methodology that integrates the CTM with GA to estimate traffic densities and calibrate the parameters of a triangular FD, leveraging vehicle trajectories extracted from camera footage to provide a practical solution for real-world traffic management.

The experiments conducted using the SUMO simulated traffic data from three different links and parameter settings validate the effectiveness of our approach. The overall result highlights the proposed method's ability to use the partially observed traffic data captured from the camera to effectively estimate density in the unobserved regions. Both the FD calibration phase and the density estimation phase, which were used on different space-time matrices across the three links used for validation, show that the proposed method can be scaled up and used in a wide range of situations.

Although the results show better performance for Links 2 and 3 compared to Link 1, this is due to the smaller dataset and limited variability in the simulated data available for Links 2 and 3. A deeper analysis was conducted to explore the effects of discretization, varying penetration rates, and traffic conditions on estimation accuracy. The CTM is particularly sensitive to discretization parameters, which significantly influence estimation results. Additionally, as penetration rates increase, estimation accuracy also improves, a trend that is particularly noticeable for Link 1, where the data spans the entire true FD. Additionally, the method's performance declines as the traffic density increases on the lane, a trend that is evident for all three links.

While the results are promising, the study's validity is constrained by several key limitations that offer clear directions for future work. Most critically, the research relies exclusively on SUMO-simulated traffic for validation, an approach that cannot fully replicate the complexity and stochastic nature of real-world conditions and may lead to significant discrepancies in practical applications. To enhance the model’s robustness, future work must therefore move beyond simulations and refine the methodology to better accommodate transient flows and complex, non-equilibrium traffic patterns present in the real world. Further exploration of advanced traffic flow models, alongside addressing other constraints like camera detection range and discretization choices, will be essential for bridging the gap between these preliminary findings and real-world efficacy.

\newpage
\bibliographystyle{apalike}
\bibliography{references}

\clearpage 
\newpage
\begin{appendices}

\newpage
\section{Density Matrix used for RMSE calculation.}\label{appendix:matrix-shape}
\begin{figure}[htbp]
    \centering
    \begin{tabular}{cc}
        \includegraphics[width=6cm]{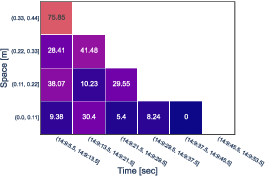} &
        \includegraphics[width=6cm]{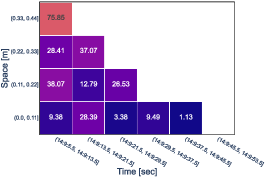} \\
        \multicolumn{1}{c}{\small (a) ground truth} & 
        \multicolumn{1}{c}{\small (b) known values} \\
    \end{tabular}
    
    \begin{tabular}{c}
        \includegraphics[width=6cm]{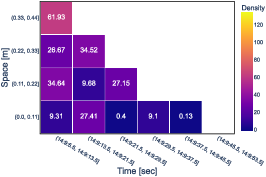} \\
        \small (c) GA estimated \\
    \end{tabular}
    
    \caption{The density matrices for the example scenario in Figure \ref{fig:example_sumo}. The matrix cells here represent only the cells located below the trajectory of the probe vehicle.}
    \label{fig:density_result}
\end{figure}

The Figure \ref{fig:density_result} shows an example of the matrices that are used for calculating RMSE in Density Estimation. We calculate the RMSE for each scenario, specifically focusing on the cells located below the trajectory of the probe vehicle. This selection is made due to the absence of information regarding the outflow in the opposite lane, making it unfeasible to predict traffic flow beyond the passage of vehicles in the opposing lane. Consequently, such cells are excluded from our analysis. In the Figure \ref{fig:density_result} shows the resultant density matrix derived from the estimated boundary conditions using GA and is compared with the ground truth density matrix and density matrix estimated with known FD values.

\newpage
\section{Runtimes for Estimation Methods}\label{appendix:runtimes}
The proposed estimation method operates offline, processing data only after a sufficient number of probe runs have been collected for a specific link. The following Table \ref{tab:fd-runtime} and Table \ref{tab:den-runtime} shows the runtimes for FD calibration and Density estimation process for all 3 links from SUMO-simulated data. These runtimes are reported for the estimation process running on Intel i5 processor using single core. The estimation methods are written in Python that uses single thread for all the sequential steps. While the individual computational steps are quick—processing one quartet for fundamental diagram estimation takes approximately $6.28 \times 10^{-5}$ seconds, and estimating density for an entire link requires about $3.29 \times 10^{-3}$ seconds—the overall runtime becomes substantial. 

\begin{table}[htbp]
\centering
\caption{FD Estimation runtime across all 3 links in SUMO experimentation.}
\label{tab:fd-runtime}
\begin{tabularx}{\textwidth}{@{}lXXX@{}}
\toprule
                           & \textbf{Link 1}   & \textbf{Link 2}   & \textbf{Link 3}   \\ \midrule
\multicolumn{4}{c}{\textbf{Amount of Data}}                                     \\ \midrule
Number of Quartet          & 4080              & 1896              & 1745              \\ \midrule
\multicolumn{4}{c}{\textbf{GA Parameters}}                                      \\ \midrule
Number of GA Solutions     & 225               & 225               & 225               \\
Number of Generations      & 100               & 100               & 100               \\ \midrule
\multicolumn{4}{c}{\textbf{Runtimes (sec)}}                                     \\ \midrule
Per Quartet                & 6,29E-05          & 6,29E-05          & 6,29E-05          \\
Per GA Generation          & 57,73             & 26,83             & 24,69             \\ \midrule
\textbf{Total Runtime}     & \textbf{5772,99}  & \textbf{2682,74}  & \textbf{2469,09}  \\ \bottomrule
\end{tabularx}
\end{table}

\begin{table}[htbp]
\centering
\caption{Density Estimation runtime across all 3 links in SUMO experimentation.}
\label{tab:den-runtime}
\begin{tabularx}{\textwidth}{@{}lXXX@{}}
\toprule
                       & \textbf{Link1}    & \textbf{Link2}    & \textbf{Link3}   \\ \midrule
\multicolumn{4}{c}{\textbf{Amount of Data}}                                       \\ \midrule
Number of Probe runs   & 1477              & 362               & 292              \\ \midrule
\multicolumn{4}{c}{\textbf{GA Parameters}}                                        \\ \midrule
Number of GA Solutions & 100               & 100               & 100              \\
Number of Generations  & 100               & 100               & 100              \\ \midrule
\multicolumn{4}{c}{\textbf{Runtimes (sec)}}                                       \\ \midrule
CTM over link          & 3,29E-03          & 3,29E-03          & 3,29E-03         \\
Per GA Generation      & 0,33              & 0,33              & 0,33             \\
Per Probe run          & 32,89             & 32,89             & 32,89            \\ \midrule
\textbf{Total Runtime} & \textbf{48572,11} & \textbf{11904,61} & \textbf{9602,61} \\ \bottomrule
\end{tabularx}
\end{table}

\end{appendices}

\end{document}